\documentclass[twocolumn,floatfix,superscriptaddress]{revtex4-2} 
\usepackage[utf8]{inputenc}
\usepackage{graphicx}
\usepackage{bigstrut}
\usepackage{amsmath}
\usepackage{comment}
\usepackage{xcolor}
\usepackage{multirow}
\usepackage{hyperref}
\hypersetup{colorlinks=true,allcolors=blue}
\usepackage{orcidlink}
\definecolor{airforceblue}{rgb}{0.36, 0.54, 0.66}

\usepackage{enumitem}

\definecolor{carminered}{rgb}{1.0, 0.0, 0.22}
\newcommand{\rev}[1]{\textcolor{black}{#1}}

\begin{document}
\title{
On the redundancy in large material datasets: efficient and robust learning with less data
}

\author{Kangming Li\orcidlink{0000-0003-4471-8527}}
\affiliation{Department of Materials Science and Engineering, University of Toronto, 27 King’s College Cir, Toronto, ON, Canada.}

\author{Daniel Persaud}
\affiliation{Department of Materials Science and Engineering, University of Toronto, 27 King’s College Cir, Toronto, ON, Canada.}

\author{Kamal Choudhary\orcidlink{0000-0001-9737-8074}}
\affiliation{Material Measurement Laboratory, National Institute of Standards and Technology, 100 Bureau Dr, Gaithersburg, MD, USA.}

\author{Brian DeCost\orcidlink{0000-0002-3459-5888}}
\affiliation{Material Measurement Laboratory, National Institute of Standards and Technology, 100 Bureau Dr, Gaithersburg, MD, USA.}

\author{Michael Greenwood}
\affiliation{Canmet MATERIALS, Natural Resources Canada, 183 Longwood Road south, Hamilton, ON, Canada.}

\author{Jason Hattrick-Simpers\orcidlink{0000-0003-2937-3188}}
\email[Correspondence: ]{jason.hattrick.simpers@utoronto.ca}
\affiliation{Department of Materials Science and Engineering, University of Toronto, 27 King’s College Cir, Toronto, ON, Canada.}

\begin{abstract}
Extensive efforts to gather materials data have largely overlooked potential data redundancy. In this study, we present evidence of a significant degree of redundancy across multiple large datasets for various material properties, by revealing that up to 95 \% of data can be safely removed from machine learning training with little impact on in-distribution prediction performance. The redundant data is related to over-represented material types and does not mitigate the severe performance degradation on out-of-distribution samples. In addition, we show that uncertainty-based active learning algorithms can construct much smaller but equally informative datasets. We discuss the effectiveness of informative data in improving prediction performance and robustness and provide insights into efficient data acquisition and machine learning training. This work challenges the ``bigger is better" mentality and calls for attention to the information richness of materials data rather than a narrow emphasis on data volume. 
\end{abstract}

\maketitle

\section{Introduction}
Data is essential to the development and application of machine learning (ML), which has now become a widely adopted tool in materials science~\cite{Butler2018,Vasudevan2019,Morgan2020a,decost2020scientific,Hart2021,Stach2021,Choudhary2022,schleder2019dft,green2022autonomous,kalinin2022machine,krenn2022scientific}. While data is generally considered to be scarce in various subfields of materials science, there are indications that the era of big data is emerging for certain crucial material properties. For instance, a substantial amount of material data has been produced through high-throughput density functional theory (DFT) calculations~\cite{horton2021promises}, leading to the curation of several large databases with energy and band gap data for millions of crystal structures~\cite{Draxl2018,AFLOWLib,Choudhary2020,Jain2013,Saal2013}.
The recently released Open Catalyst datasets contain over 260 million DFT data points for catalyst modeling~\cite{Chanussot2021,Tran2022a}. The quantity of available materials data is expected to grow at an accelerated rate, driven by the community's growing interest in data collection and sharing.

In contrast to the extensive effort to gather ever larger volume of data, information richness of data has so far attracted little attention. Such a discussion is important as it can provide critical feedback to data acquisition strategies adopted in the community. For instance, DFT databases were typically constructed either from exhaustive enumerations over possible chemical combinations and known structural prototypes or from random sub-sampling of such enumerations~\cite{AFLOWLib,Jain2013,Saal2013,Choudhary2020,Chanussot2021,Tran2022a,Shen2022}, but the effectiveness of these strategies in exploring the materials space remains unclear. Furthermore, existing datasets are often used as the starting point for the data acquisition in the next stage. For example, slab structures in Open Catalyst datasets were created based on the bulk materials from Materials Project~\cite{Chanussot2021,Tran2022a}. Redundancy in the existing datasets, left unrecognized, may thus be passed on to future datasets, making subsequent data acquisition less efficient.

In addition, examining and eliminating redundancy in existing datasets can improve training efficiency of ML models. Indeed, the large volume of data already presents significant challenges in developing ML models due to the increasingly strong demand for compute power and long training time. For example, over 16,000 GPU days were recently used for analyzing and developing models on the Open Catalyst datasets~\cite{Gasteiger2022}. Such training budgets are not available to most researchers, hence often limiting model development to smaller datasets or a portion of the available data~\cite{Choudhary2023}. On the other hand, recent work on image classification has shown that a small subset of data can be sufficient to train a model with performance comparable to that obtained using the entire dataset~\cite{Yang2022,Sorscher2022}. It has been reported that aggressively filtering training data can even lead to modest performance improvements on natural language tasks, in contrast to the prevailing wisdom of ``bigger is better'' in this field~\cite{Geiping2022}. To the best of our knowledge, however, there has been no investigation of the presence and degree of data redundancy in materials science. Revealing data redundancy can inform and motivate the community to create smaller benchmark datasets, hence significantly scaling down the training costs and facilitating model development and selection. \rev{This may be important in the future if data volume grows much faster than the available training budget, which is a likely scenario, as data volume is proportional to resources available to the entire community, while training budgets are confined to individual research groups.}

The examination of data redundancy is also important in other scenarios in materials science. Methods developed for selecting the most informative data can be used as the strong baselines for active learning algorithms, which are increasingly common in ML-driven materials discovery workflows~\cite{Ling2017,Smith2018,Lookman2019,jia2019anthropogenic,Zhong2020,Kusne2020,rohr2020benchmarking,liang2021benchmarking,Wang2022c}. Analysis of information richness can also improve our understanding of the material representation and guide the design of active learning algorithms. In the multi-fidelity data acquisition setting~\cite{Kingsbury2022}, one can perform high-fidelity measurement only on the informative materials down-selected from the larger but low-fidelity datasets.

In this work we present a systematic investigation of data redundancy across multiple large material datasets by examining the performance degradation as a function of training set size for traditional descriptor-based models and state-of-the-art neural networks. To identify informative training data, we propose a pruning algorithm and demonstrate that smaller training sets can be used without substantially compromising the ML model performance, highlighting the issue of data redundancy.
We also find that selected sets of informative materials transfer well between different ML architectures, but may transfer poorly between substantially different material properties. Finally, we compare uncertainty-based active learning strategies with our pruning algorithm, and discuss the effectiveness of active learning for more efficient high throughput materials discovery and design.

\section{Results}
\subsection{Redundancy evaluation tasks}
We investigate data redundancy by examining the performance of ML models. To do so, we use the standard hold-out method for evaluating ML model performance: We create the training set and the hold-out test set from a random split of the given dataset. The training set is used for model training, while the test set is reserved for evaluating the model performance. In the following, we refer to the performance evaluated on this test set as the in-distribution (ID) performance, and this training set as the pool. To reveal data redundancy, we train a ML model on a portion of the pool and check whether its ID performance is comparable to the one resulting from using the entire pool. Since ID performance alone may not be sufficient to prove the redundancy of the remaining unused pool data, we further evaluate the prediction performance on the unused pool data and out-of-distribution (OOD) test data. 

\begin{figure}[btp]
\centering
	\includegraphics[width=\linewidth]{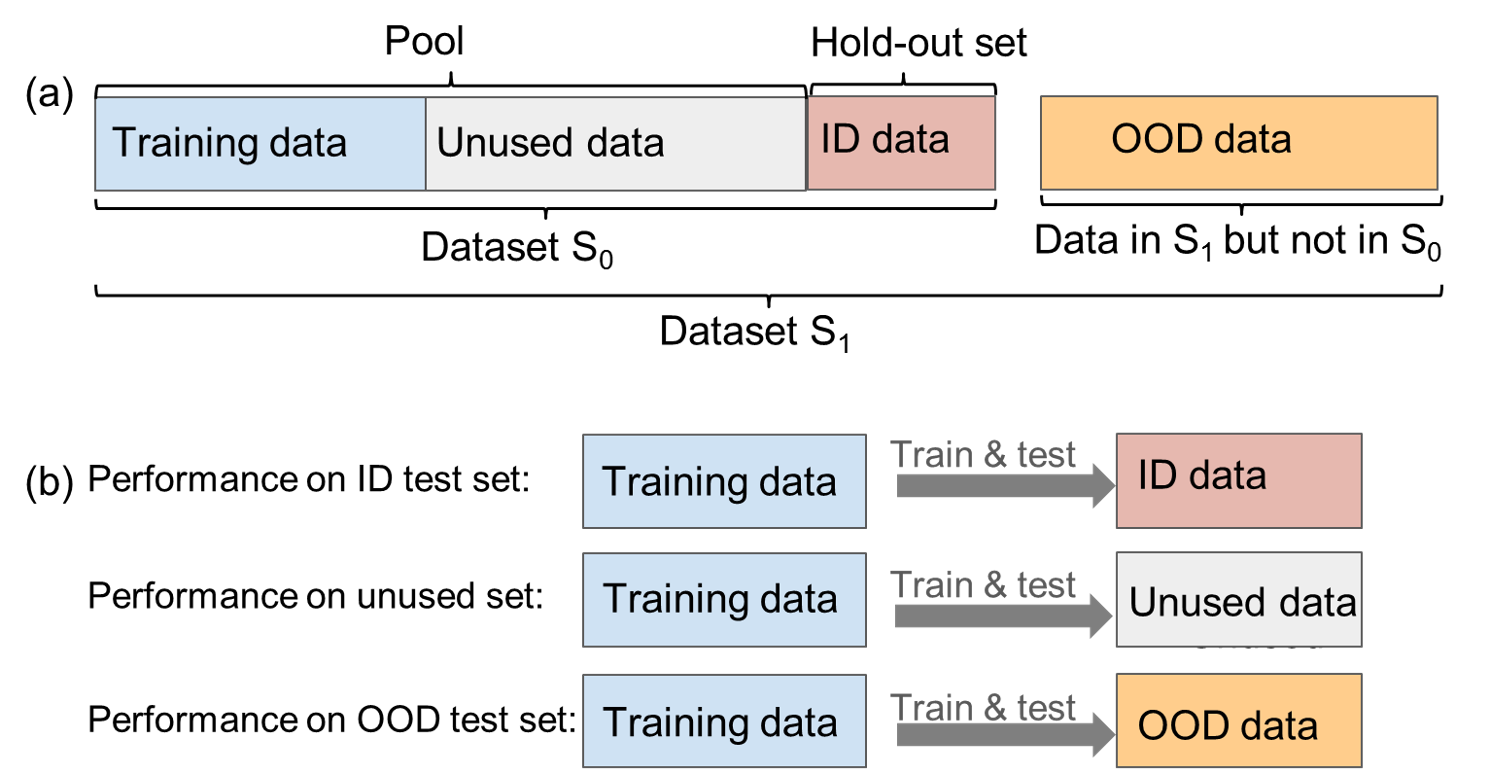}
\caption{\label{fig:schema} \textbf{Schematic of redundancy evaluation}. \textbf{a} the dataset splits. \textbf{b} three prediction tasks to evaluate model performance and data redundancy. }
\end{figure}

Fig.\ref{fig:schema} illustrates the redundancy evaluation discussed above. We first perform a (90, 10)~\% random split of the given dataset $S_0$ to create the pool and the ID test set. To create an OOD test set, we consider new materials included in a more recent version of the database $S_1$. Such OOD sets enable the examination of model performance robustness against distribution shifts that may occur when mission-driven research programs focus on new areas of material space~\cite{li2022critical}. We progressively reduce the training set size from 100~\% to 5~\% of the pool via a pruning algorithm (see Methods). ML models are trained for each training set size, and their performance is tested on the hold-out ID test data, the unused pool data, and the OOD data, respectively.

To ensure a comprehensive and robust assessment of data redundancy, we examine the formation energy, band gap, and bulk modulus data in three widely-used DFT databases, namely JARVIS~\cite{Choudhary2020}, Materials Project (MP)~\cite{Jain2013}, and OQMD~\cite{Saal2013}. For each database, we consider two release versions to study the OOD performance and to compare the data redundancy between different database versions. The number of entries for these datasets is given in Table~\ref{tab:datasets}. 

To ascertain whether data redundancy is model-agnostic, we consider two conventional ML models, namely XGBoost (XGB)~\cite{xgboost} and random forests (RF)~\cite{breiman2001random}, and a graph neural network called the Atomistic LIne Graph Neural Network (ALIGNN)~\cite{Choudhary2021}. The RF and XGB models are chosen since they are among the most powerful descriptor-based algorithms~\cite{Zhang2023}, whereas ALIGNN is chosen as the representative neural network because of its state-of-the-art performance in the Matbench test suite~\cite{Dunn2020} at the time of writing. 

\begin{table}[btp]
  \centering
  \caption{Number of entries of formation energy ($E_f$), band gap ($E_g$), and bulk modulus ($K$) data in different datasets. The last two digits in the dataset name indicate the year of release (e.g. MP18 for the 2018 version). 
  }
  \label{tab:datasets}
\begin{ruledtabular}
\begin{tabular}{ccccccc}
 &JARVIS18 & JARVIS22 & MP18 & MP21 & OQMD14 & OQMD21 \\ \hline
$E_f$ & 53k      & 76k      & 68k  & 146k & 290k   & 1M     \\
$E_g$ & 53k      & 76k      & 68k  & 146k & 290k   & 1M     \\
$K$ & 19k   & 24k      & 7k   & 7k   & 0      & 0     
\end{tabular}
\end{ruledtabular}
\end{table}

\subsection{In-distribution performance\label{sec:ID_performance}}

We begin by presenting an overview of the ID performance for all the model-property-dataset combinations in Table~\ref{tab:ID_RMSE}, where the root mean square errors (RMSE) of the models trained on the entire pool are compared to those obtained with 20~\% of the pool. For brevity, we refer to the models trained on the entire pool and on the subsets of the pool as the full and reduced models, respectively, but we note that the model specification is the same for both full and reduced models and the terms ``reduced" and ``full" pertain only to the amount of training data.

For the formation energy prediction, the RMSE of the reduced RF models increase by less than 6~\% compared to those of the full RF models in all cases. Similarly, the RMSE of the reduced XGB models increase only by 10~\% to 15~\% compared to the RMSE of the full XGB models in most datasets, except in OQMD21 where a 3~\% decrease in the RMSE is observed. The RMSE of the reduced ALIGNN models increase by 15~\% to 45~\%, a larger increment than observed for the RF and XGB models. Similar trend is observed for the band gap and bulk modulus prediction, where the RMSE of the reduced models typically increase by no more than 30~\% compared to those of the full models.

\begin{table}[tbp]
  \centering
  \caption{RMSE scores on the ID test sets using the full and reduced models. The standard deviation (STD) of labels is also given in the second column. The reduced models are trained on the subset (20 \% of the pool) selected via the pruning algorithm. The ALIGNN results for the formation energy and band gap data in OQMD21 are not available because of the high training cost associated with the large data volume. 
  }
  \label{tab:ID_RMSE}
  \begin{ruledtabular}
\begin{tabular}{cccccccc}
\multirow{2}{*}{Dataset} & \multirow{2}{*}{STD} & \multicolumn{2}{c}{RF} & \multicolumn{2}{c}{XGB} & \multicolumn{2}{c}{ALIGNN} \\
\cline{3-8}
         &      & Full  & 20 \% & Full  & 20 \% & Full  & 20 \% \\ \hline
         & \multicolumn{7}{c}{Formation energy  (eV/atom)}         \\
JARVIS18 & 1.08 & 0.187 & 0.190   & 0.136 & 0.159   & 0.064 & 0.093   \\
JARVIS22 & 1.08 & 0.191 & 0.196   & 0.149 & 0.165   & 0.074 & 0.102   \\
MP18     & 1.06 & 0.159 & 0.168   & 0.120 & 0.140   & 0.065 & 0.085   \\
MP21     & 1.21 & 0.190 & 0.196   & 0.161 & 0.175   & 0.081 & 0.093   \\
OQMD14   & 0.85 & 0.117 & 0.124   & 0.096 & 0.105   & 0.058 & 0.068   \\
OQMD21   & 1.00 & 0.117 & 0.123   & 0.109 & 0.104   & /     & /       \\ \hline
          & \multicolumn{7}{c}{Band gap  (eV)}                      \\ 
JARVIS18 & 1.41 & 0.433 & 0.506   & 0.404 & 0.439   & 0.395 & 0.497   \\
JARVIS22 & 1.33 & 0.406 & 0.465   & 0.385 & 0.411   & 0.365 & 0.441   \\
MP18     & 1.62 & 0.613 & 0.738   & 0.587 & 0.658   & 0.613 & 0.743   \\
MP21     & 1.51 & 0.555 & 0.683   & 0.535 & 0.616   & 0.529 & 0.682   \\
OQMD14   & 0.72 & 0.211 & 0.212   & 0.196 & 0.198   & 0.185 & 0.189   \\
OQMD21   & 0.87 & 0.308 & 0.314   & 0.314 & 0.323   & /     & /       \\ \hline
          & \multicolumn{7}{c}{Bulk modulus (GPa)}                  \\ 
JARVIS18 & 66.6 & 24.6  & 26.8    & 23.7  & 27.1    & 22.9  & 29.6    \\
MP18     & 75.8 & 22.0  & 23.0    & 18.7  & 24.2    & 16.0  & 31.2   
\end{tabular}
\end{ruledtabular}
\end{table}

\begin{figure*}[htbp]
\centering
\includegraphics[width=\linewidth]{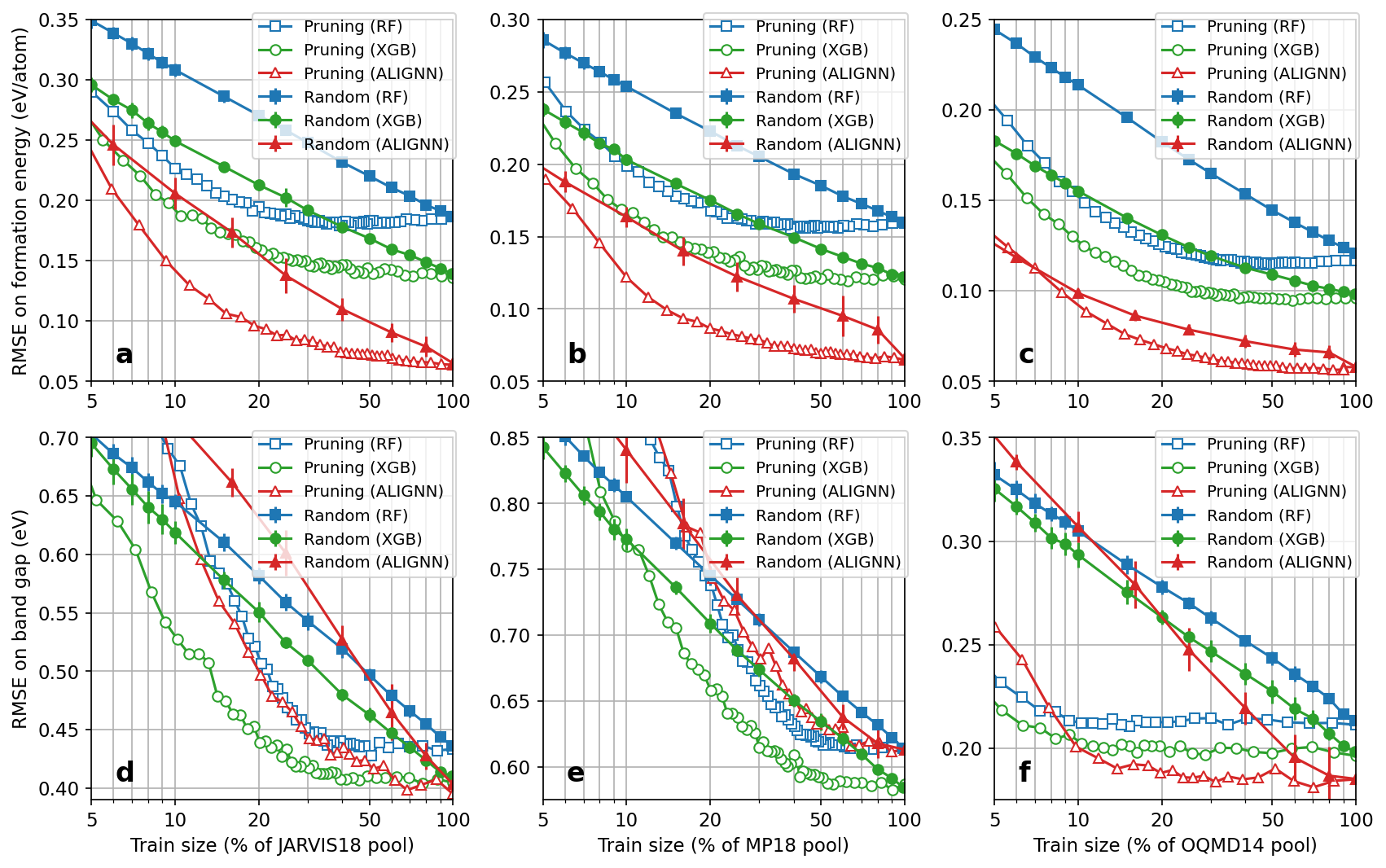}
\caption{\label{fig:ID performance}\textbf{RMSE on the ID test sets.} \textbf{a-c} JARVIS18, MP18, and OQMD14 formation energy prediction. \textbf{d-f} JARVIS18, MP18, and OQMD14 band gap prediction. The random baseline results to for the XGB and RF (or ALIGNN) models are obtained by averaging over the results of 10 (or 5) random data selections for each training set size. The $X$ axis is in the log scale.}
\end{figure*}

Next, we conduct a detailed analysis for formation energy and band gap properties because of their fundamental importance for a wide range of materials design problems. Fig.~\ref{fig:ID performance} shows the ID performance as a function of training set size (in percentage of the pool) for the formation energy and band gap prediction in the JARVIS18, MP18 and OQMD14 datasets. Results for other datasets can be found in Supplementary Figure 1-6.

For the formation energy prediction, the prediction error obtained with the pruned data drops much faster with increasing data size than the one obtained using the randomly selected data. When accounting for more than 5~\% of the training pool, the pruned datasets lead to better ID performance than the ones from random sampling. In particular, the RF, XGB, and ALIGNN models trained with 20~\% of the pool selected by the pruning algorithm have the same ID performance as the ones trained with a random selection of around 90~\%, 70~\%, and 50~\%, respectively, of the pool. 

A large portion of training data can be removed without significantly hurting the model performance. To demonstrate this, we define a quantitative threshold for the ``significance'' of the performance degradation as a 10~\% relative increase in RMSE; data that can be pruned without exceeding this performance degradation threshold are considered redundant. With this definition, only 13~\% of the JARVIS18 data, and 17~\% of the MP18 and OQMD data are informative for the RF models. For the XGB models, between 20~\% and 30~\% of the data are needed depending on the datasets. For the ALIGNN models, 55~\%, 40~\% and 30~\% of the JARVIS18, MP18 and OQMD14 data are informative, respectively. While the JARVIS18 dataset may seem to be less redundant for the ALIGNN models, \rev{the 10~\% increase in the RMSE (60 meV/atom) corresponds to an RMSE increase of only 6 meV/atom}, much smaller than the DFT accuracy of around 100 meV/atom with respect to experiments~\cite{Kirklin2015}. In fact, training the ALIGNN model on 30~\% of the JARVIS18 data only leads to a drop of 0.002 in the $R^2$ test score.

\rev{
While this work is focused on redundancy which is model and dataset specific, it is still worth commenting on the model performance scaling across models and datasets. When using the random sampling for data selection, we observe a power law scaling for all the models and datasets. For formation energy datasets, switching the models mainly shifts the scaling curve without much change to the slopes. For band gap datasets, switching from RF to XGB models shifts the scaling curve down without changing the slope, whereas switching from tree-based models to ALIGNN leads to a steeper slope and hence better scaling. Compared to training on randomly sampled data, training on informative data as selected by the pruning algorithm can lead to better scaling until reaching saturation when there is no more informative data in the pool. Different datasets exhibit similar scaling behaviors with the slope and saturation point dependent on target property and material space covered by the datasets.
}

The performance response to the size of band gap data is similar to that observed in the formation energy data. The redundancy issue is also evident in band gap data: a 10~\% RMSE increase corresponds to training with 25~\% to 40~\% of the data in the JARVIS18 and MP18 datasets. Even more strikingly, only 5~\% (or 10~\%) of the OQMD14 band gap data are sufficiently informative for the RF and XGB (or ALIGNN) models. 

\begin{figure}[!htbp]
\centering
\includegraphics[width=\linewidth]{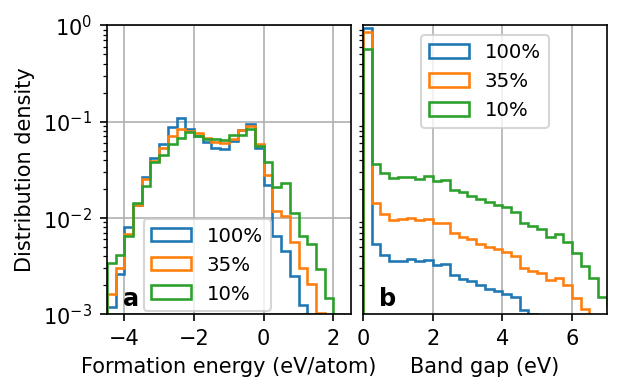}
\caption{\label{fig:label dist}
\rev{Label distributions of the XGB-pruned training sets. \textbf{a} MP18 formation energy data. \textbf{b} OQMD14 band gap data. The legend indicates the training set size in percentage of the pool. Results for other datasets can be found in Supplementary Figure 15 and 16.}
}
\end{figure}

These results demonstrate the feasibility of training on only a small portion of the available data without much performance degradation. We find that this is achieved by skewing the data distribution towards the underrepresented materials. For instance, the distributions of the pruned data are skewed towards materials with large formation energies and band gaps \rev{(Fig.~\ref{fig:label dist})}, which are both underrepresented and less accurately predicted materials. These results not only confirm the importance of the data diversity~\cite{Zhang2023} but also highlight the redundancy associated with overrepresented materials.

ID performance is not sufficient to prove that the unused data are truly redundant. The effects related to model capability and the test set distribution should also be considered. Indeed, one may argue that the current ML models (in particular, the band gap models) are not advanced enough to learn from the unused data leading to a false sense of the data redundancy. Furthermore, the similar performance of the full and reduced models does not imply a similar performance on a test set following a different distribution. These questions are addressed in the following two sections by discussing the performance on the unused data and on the OOD data. 

\subsection{Performance on unused data\label{sec:redundant}}

\begin{figure*}[!htbp]
\centering
\includegraphics[width=\linewidth]{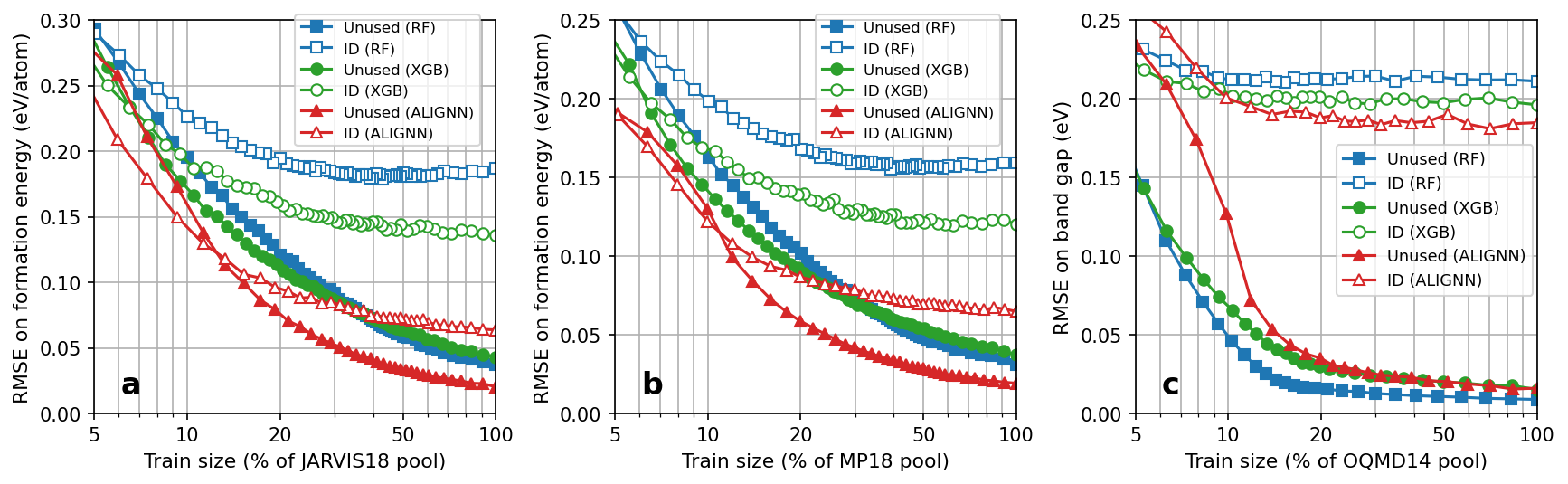}
\caption{\label{fig:redundant performance}\textbf{RMSE on the unused data in the pool.} \textbf{a} JARVIS18 formation energy prediction. \textbf{b} MP18 formation energy prediction. \textbf{c} OQMD14 band gap prediction. Performance on the ID test set is shown for comparison.}
\end{figure*}

Here we further examine the model performance on the unused pool data. Fig.~\ref{fig:redundant performance} shows three representative cases: the JARVIS18 and MP18 formation energy datasets, and the OQMD14 band gap dataset. 
For the formation energy prediction, the RMSE on the unused data become lower than on the ID RMSE when the training set size is above 5~\% to 12~\% of the pool, and is half of the ID RMSE when the training set size is above 30~\% to 40~\% of the pool. Similar trend is observed for the band gap prediction with varying thresholds of the performance improvement saturation depending the datasets (Supplementary Figure 10-12). In particular, the OQMD14 results in Fig.~\ref{fig:redundant performance} show that the models trained on 10~\% of the pool can well predict the unused data that account for 90~\% of the pool, with the associated RMSE much lower than the RMSE on the ID test set. The good prediction on the unused data signifies a lack of new information in these data, confirming that the improvement saturation in the ID performance is caused by the information redundancy in the unused data rather than the incapability of models to learn new information.

\rev{
While the scaling curve for the unused data has a shape similar to the one for the ID test data, the former shows a much steeper slope for the training set sizes below 15\% of the pool, and reaches saturation at a slower rate. In addition, it is noted that the ranking of different ML models for their performance on the unused data is not necessarily the same as for the ID test data. For instance, for the JARVIS18 and MP18 formation energy data, the XGB model outperforms the RF model on the ID test set whereas their performance is practically the same on the unused data. Among the models trained on the OQMD14 band gap data, the RF model has the largest RMSE on the ID test set but the lowest error on the unused data.
}
\subsection{Out-of-distribution performance}

To check whether redundancy in training data also manifests under a distribution shift in test data, we examine the model performance on the OOD test data consisting of the new materials in the latest database versions (JARVIS22, MP21, and OQMD21) using the models trained on the older versions (JARVIS18, MP18 and OQMD14). 

\begin{figure*}[htbp]
\centering
\includegraphics[width=\linewidth]{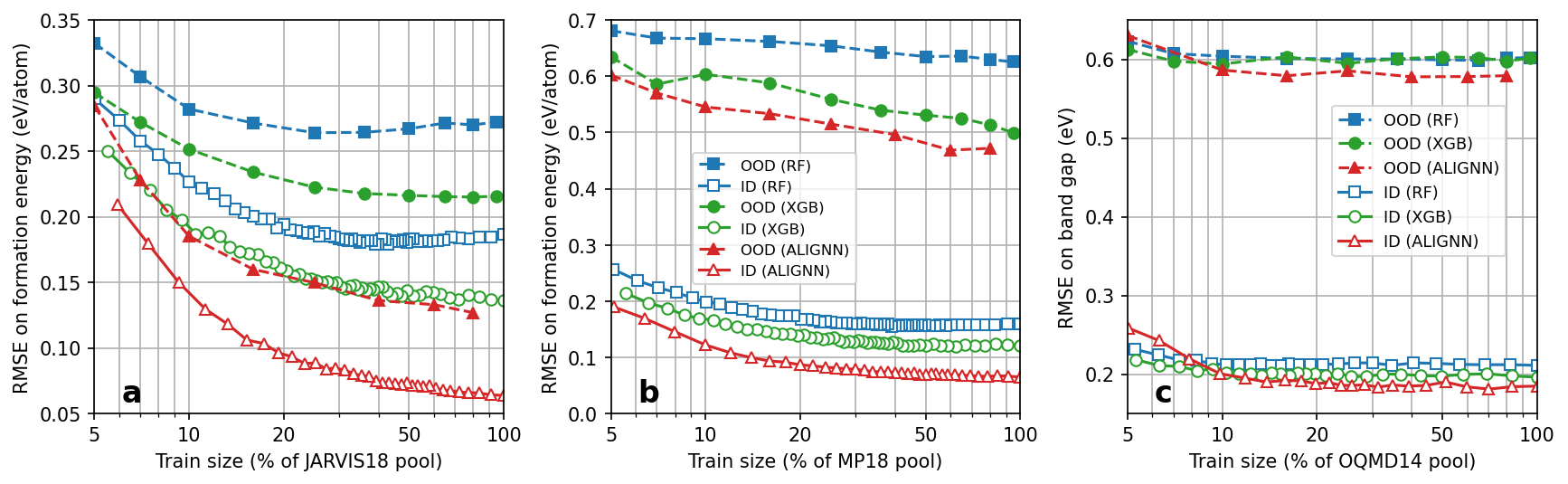}
\caption{\label{fig:ood performance}\textbf{RMSE on the OOD test sets.} \textbf{a} JARVIS formation energy prediction. \textbf{b} MP formation energy prediction. \textbf{c} OQMD band gap prediction. Performance on the ID test set is shown for comparison. 
\rev{The reader interested in the statistical overlaps between the ID and OOD data in the feature space is referred to Supplementary Figure 24.}
}
\end{figure*}

First, we find that training on the pruned data can lead to better or similar OOD performance than the randomly sampled data of the same size. We therefore focus here on the OOD performance based on the pruned data shown in Fig.~\ref{fig:ood performance}. \rev{
Overall, the scaling curves for the OOD performance show are similar to those for the ID performance with slightly different slopes and saturation data size, }confirming the existence of the data redundancy measured by the OOD performance. Specifically, using 20~\%, 30~\%, or 5~\% to 10~\% of the JARVIS18, MP18, or OQMD14 data, respectively, can lead to an OOD performance similar to that of the full models, with around 10~\% RMSE increase. 

The performance on OOD data can be severely degraded. Even for the models trained on the entire pool, the increase in the OOD RMSE with respect to the ID RMSE often goes above 200~\% for the considered databases and can rise up to 640~\% in the case of the ALIGNN-MP formation energy prediction (Supplementary Table 1). Therefore, the excellent ID performance obtained with state-of-the-art models and large datasets might be a catastrophically optimistic estimation of the true generalization performance in a realistic materials discovery setting~\cite{li2022critical,Zhang2023}. 

\rev{
Different databases exhibit a varying degree of performance degradation, which should be correlated with the degree of statistical overlaps between the database versions rather than the quality of the databases. In fact, database updates that induce such performance degradation are desirable because they are indications of new ``unknown" observations and can lead to more robust generalization performance. One interesting line of research would be therefore to develop methods to deliberately search for materials where the previous models would fail catastrophically as a path to expand a database.
}


\rev{
The strong OOD performance degradation highlights the importance of information richness over data volume. It also raises an interesting question: given a training set $A_1$, is it possible to find a smaller training set $A_2$ such that the $A_2$-trained model perform similarly to the $A_1$-trained model on an $A_1$-favorable test set $B_1$ (i.e., same distribution as $A_1$) but significantly outperform the $A_1$-trained model on an $A_1$-unfavorable test set $B_*$ (i.e., distribution different from $A_1$)?} 
Indeed, we find that training on the heavily pruned MP21 pool ($A_2$) gives dramatically better prediction performance on the MP21 test data ($B_*$) than training on 10$\times$ more data from the MP18 pool ($A_1$) whereas their performance is similar on the MP18 test set ($B_1$). \rev{The result confirms the idea of finding a training set whose distribution can not only well cover but also significantly extend beyond the original one while still being much smaller in size. The result
highlights that information richness and data volume are not necessarily correlated, and the former is much more important for the prediction robustness.} By covering more materials within the data distribution, we may better ensure unknown materials are from known distributions (``known unknown") and avoid unexpected performance degradation (``unknown unknown"), which is particularly important in scenarios such as materials discovery or building universal interatomic potentials~\cite{Takamoto2022,Chen2022,Choudhary2023}.

\subsection{Transferability of pruned material sets}
The ID performance results demonstrate that our pruning algorithm effectively identifies informative material sets for a given ML model and material property. A natural followup inquiry is the universality, or more specifically, the transferability of these sets between ML architectures and material properties. 

We find a reasonable level of transferability of the pruned material set across ML architectures, confirming that data pruned by a given ML architecture remains informative to other ones (Supplementary Figure 17-20). For example, XGB models trained on RF-pruned data outperform those trained on twice as much randomly selected data for formation energy prediction. Moreover, the XGB model still outperforms an RF model trained on the same pruned data, consistent with our observed performance ranking (XGB$>$RF). This ensures robustness against information loss with respect to future architecture change: more capable models developed in the future can be expected to extract no less information from the pruned dataset than the current state-of-the-art one, even if the dataset is pruned by the latter. It would therefore be desirable to propose benchmark datasets pruned from existing large databases using current models, which can help accelerate the development of ML models due to the smaller training cost.

In contrast, we find that there is a limited transferability of pruned datasets across different material properties. For instance, the band gap models trained on the pruned formation energy data outperform those trained on randomly sampled data by only by a slight margin (Supplementary Figure 21), suggesting little overlap between informative material sets for predicting these two properties. This limited task transferability may be a result of the lack of strong correlation between the formation energy and band gap data, for which the Spearman correlation coefficient is -0.5 in the considered databases. Additionally, the OOD results show that formation energy and band gap models do not necessarily suffer the same degree of performance degradation when tested on new materials despite being trained on the same set of materials (Supplementary Table 1), indicating learned feature-property relations could differ significantly. These considerations suggest that a fruitful line of future research might explore dataset pruning based on multitask regression models focusing on a diverse set of material properties controlled by different underlying physical phenomena.

\subsection{Uncertainty-based active learning}

\begin{figure*}[htbp]
\centering
\includegraphics[width=\linewidth]{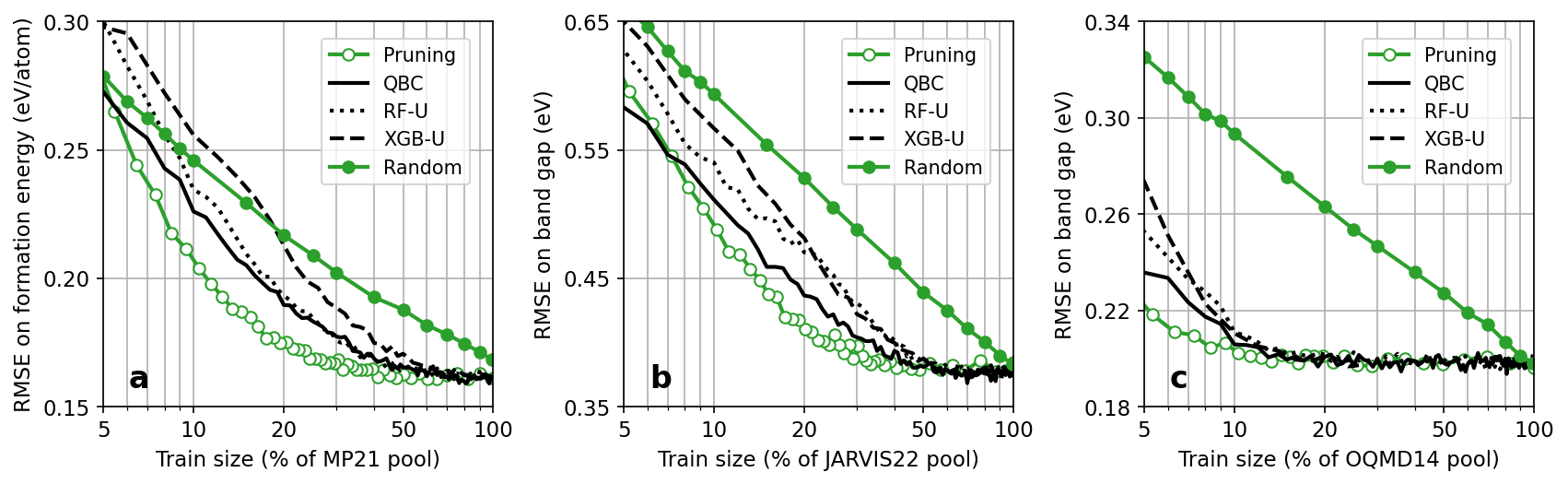}
\caption{\label{fig:AL IID} \textbf{RMSE on the ID test sets by the XGB models trained on the data selected using the active learning algorithms.} \textbf{a} MP21 formation energy prediction. \textbf{b} JARVIS22 formation energy prediction. \textbf{c} OQMD14 band gap prediction. QBC: query by committee, RF-U: random forest uncertainty, XGB-U: XGBoost uncertainty. The performance obtained using the random sampling and the pruning algorithm is shown for comparison.
}
\end{figure*}

In the previous sections we have revealed the data redundancy in the existing large material databases through dataset pruning. How much, then, can we avoid such data redundancy in the first place when constructing the databases? To this end, we consider active learning algorithms that select samples with largest prediction uncertainty (see Methods). The first and the second algorithms use the width of the 90~\% prediction intervals of the RF and XGB models as the uncertainty measure, respectively, whereas the third one is based on the query by committee (QBC), where the uncertainty is taken as the disagreement between the RF and XGB predictions. 

Fig.~\ref{fig:AL IID} shows a comparison of the ID performance of the XGB models trained on the data selected using the active learning algorithm, the pruning algorithm and the random sampling. The QBC algorithm is found to be the best performing active learning algorithm. For the formation energy prediction across the three databases, 30~\% to 35~\% of the pool data selected by the QBC algorithm is enough to achieve the same model performance obtained with 20~\% of the pool data using the pruning algorithm. Furthermore, the resulting model performance is equivalent to that obtained with 70~\% to 90~\% of the pool using the random sampling. As for the band gap prediction, the models trained on the QBC-selected data perform similarly to those trained on the pruned data, or even sometimes outperform the latter when the data volume is below 20~\% (Supplementary Figure 23). In particular, the QBC algorithm can effectively identify 10~\% of the OQMD14 band gap data as the training data without hurting the model performance (Fig.~\ref{fig:AL IID}\textbf{c}).
Similar trends are also found for the RF models and for other datasets (Supplementary Figure 23).

Overall, our results across multiple datasets suggest that it is possible to leverage active learning algorithms to query only 30~\% of the existing data with a relatively small accuracy loss in the ID prediction. The remaining 70~\% of the compute may then be used to obtain a larger and more representative material space. Considering the potentially severe performance degradation on OOD samples which are likely to be encountered in material discovery, the gain in the robustness of ML models may be preferred over the incremental gain in the ID performance. 

\section{Discussion}

\rev{It is worth emphasizing that this work is by no means critical of the curation efforts or significance of these materials datasets.  Indeed, many datasets were not originally generated for ML training but as the results of long-term project-driven computational campaigns. Some of them were even curated before the widespread use of ML and have played a significant role in fueling the fast application of ML in materials science. On the other hand, the presence and degree of redundancy in a dataset is worth discussing irrespective of the original purpose. Furthermore, ML should be considered not only as a purpose, though it has become the primary use case of these datasets, but also as a statistical means or data-science tool to examine these datasets.}

This work is also not to oppose the use of big data, but to advocate a critical assessment of the information richness in data, which has been largely overlooked due to a narrow emphasis on data volume. As materials science transitions towards a big data-driven approach, such evaluations and reflections on current practices and data can offer insights into more efficient data acquisition and sensible resource usage. For instance, conventional high-throughput DFT often relies on enumerations over structural prototypes and chemical combinations. The substantial redundancy revealed in this work suggests these strategies are suboptimal in querying new informative data, whereas uncertainty based active learning can enable a $3\times$ to $10\times$ boost in  sampling efficiency. Our scaling results for OOD performance degradation further highlight the importance of information richness over sheer volume for robust predictive models. In this regard, it is preferable to allocate more resources to explore a diverse materials space rather than seeking incremental improvements in prediction accuracy within limited or well-studied regions. This may represent a paradigm shift from systematic high-throughput studies, where we can start with uncertainty based active learning in a much larger design space, and then reconsider the design space by interrogating the model and switching to a property optimization objective.

\rev{While the pruning algorithm is proposed here to illustrate data redundancy, such data selection algorithms can have other use cases, e.g., inform the design of active learning algorithms.} Indeed, the observation that data redundancy predominantly involves overrepresented materials implies that information entropy might also serve as a promising criterion for data acquisition~\cite{hennig2012entropy,Zhang2023}. A detailed analysis of pruned material sets may also offer insights into material prototypes and improve understanding of feature-property relationships,\rev{including identifying specific groups of redundant materials as well as identifying patterns that explain the poor task transferability of pruned datasets. }
Finally, the pruning algorithm offers a new funneling strategy for prioritizing materials for high-fidelity measurements. For instance, \rev{pruning the existing DFT data obtained with generalized gradient approximation (GGA) functionals can point to the materials to be recomputed with high-fidelity meta-GGA functionals~\cite{Kingsbury2022}.}


\rev{
We demonstrate that transferability of compact datasets is reasonable across models but is limited across tasks (materials properties). It is discussed in the context of data pruning, but the idea and implication hold for active learning. The limited task transferability indicates that the maximally compact set of materials for property A is not ensured to be the maximally compact set for property B. While this is an interesting observation and invites further investigation, it is not a practical issue for active learning when the measurements of two properties are independent. For example, DFT calculations of band gap and elastic modulus are unrelated, therefore the maximally compact sets of materials can be constructed independently via active learning and need not be the same. For correlated property measurements, however, more careful planning is required. For instance, the calculations of more ``expensive" properties such as band gap and elastic modulus would also give the formation energy of the same material since energy is a basic output of any DFT calculations. While the compact datasets for band gap and elastic modulus can still be searched independently without considering formation energy data, the construction of the compact dataset for formation energy should consider the data that can be obtained as by-products from the band gap and elastic modulus calculations. 
}

In conclusion, we investigate data redundancy across multiple material datasets using both conventional ML models and state-of-the-art neural networks. We propose a pruning algorithm to remove uninformative data from the training set, resulting in models that outperform those trained on randomly selected data of the same size. Depending on the dataset and ML architecture, up to 95~\% of data can be pruned with little degradation in in-distribution performance (defined as $<10~\%$ increase in RMSE) compared to training on all available data. The removed data, mainly associated with over-represented material types, are shown to be well predicted by the reduced models trained without them, confirming again the information redundancy. Using new materials in newer database versions as the out-of-distribution test set, we find that 70~\% to 95~\% of data can be removed from the training set without exceeding a 10~\% performance degradation threshold on out-of-distribution data, confirming again that the removed data are redundant and do not lead to improved performance robustness against distribution shift. Transferability analysis shows that the information content of pruned datasets transfers well to different ML architectures but less so between material properties. Finally, we show that the QBC active learning algorithm can achieve an efficiency comparable to the pruning algorithm in terms of finding informative data, hence demonstrating the feasibility of constructing much smaller material databases while still maintaining a high level of information richness.

\section*{Methods}

\subsection*{Materials datasets}

The 2018.06.01 version of Materials Project (MP18), and the 2018.07.07 and 2022.12.12 versions of JARVIS (JARVIS18 and JARVIS22) were retrieved by using JARVIS-tools~\cite{Choudhary2020}. The 2021.11.10 version of Materials Project (MP21) was retrieved using the Materials Project API~\cite{Jain2013}. The OQMD14 and OQMD21 data were retrieved from \url{https://oqmd.org/download}.

The JARVIS22, MP21, OQMD21 data were preprocessed as follows. First, entries of materials with a formation energy larger than 5 eV/atom were removed. Then, the Voronoi tessellation scheme~\cite{Ward2017a} as implemented in Matminer~\cite{Ward2018} were used to extract 273 compositional and structural features. The Voronoi tessellation did not work for a very small number of materials and these materials were removed. 

For the older versions (JARVIS18, MP18, OQMD14), we did not directly use the structures and label values from the older database. Instead, we use the materials identifiers from the older database to search for the corresponding structures and label values in the newer database. This is to avoid potential inconsistency caused by the database update.

\subsection*{ML models}
We considered three ML models here: XGB~\cite{xgboost}, RF~\cite{breiman2001random}, and a graph neural network called the Atomistic LIne Graph Neural Network (ALIGNN)~\cite{Choudhary2021}. XGB is a gradient-boosted method that builds sequentially a number of decision trees in a way such that each subsequent tree tries to reduce the residuals of the previous one. RF is an ensemble learning method that combines multiple independently built decision trees to improve accuracy and minimize variance. ALIGNN constructs and utilizes graphs of interatomic bonds and bond angles. 

We used the RF model as implemented in the scikit-learn 1.2.0 package~\cite{sklearn}, and the XGB model as implemented in the XGBoost 1.7.1 package~\cite{xgboost}. For the RF model, we used 100 estimators, 30~\% of the features for the best splitting, and default settings for other hyperparameters. We used a boosted random forest for the XGB model: 4 parallel boosted trees were used; for each tree, we used 1000 estimators, a learning rate of 0.1, an L1 (L2) regularization strength of 0.01 (0.1), and the histogram tree grow method; we set the subsample ratio of training instances to 0.85, the subsample ratio of columns to 0.3 when constructing each tree, and the subsample ratio of columns to 0.5 for each level. The hyperparameter set was kept to be the same in all the model training for the following reasons: First, performing hyperparameter tuning every time when changing the size of the training set would be very computationally expensive. Second, we verified that the model performance using the optimal hyperparameters tuned from the randomized cross-validation search was close to the one using the chosen hyperparameters. 

For the ALIGNN model, we used 2 ALIGNN layers, 2 GCN layers, a batch size of 128, and the layer normalization, while keeping other hyperparameters the same as in the original ALIGNN implementation~\cite{Choudhary2021}. We trained the ALIGNN model for 50 epochs as we found more epochs did not lead to further performance improvement. We used the same OneCycle learning rate schedule, with 30~\% of the training budget allocated to linear warmup and 70~\% to cosine annealing.

\subsection*{Pruning algorithm}
We proposed a pruning algorithm that starts with the full training pool and iteratively reduces the training set size. 
\rev{
We denote the full training pool as $D_{\rm pool}$, the training set at the $i$-th iteration as $D_{\rm train}^{i}$, the unused set as $D_{\rm unused}^{i}$ ($=D_{\rm pool} - D_{\rm train}^{i}$), and the trained model as $M^i$. At the initial iteration ($i=0$), $D_{\rm train}^{0} = D_{\rm pool}$ and $D_{\rm unused}^{0}$ is empty. At each iteration $i>0$, $D_{\rm train}^{i}$ and $D_{\rm unused}^{i}$ are updated as follows: First, a random splitting of $D_{\rm train}^{i-1}$ is performed to obtained two subsets $D_A^i$ (80 \% of $D_{\rm train}^{i-1}$) and $D_B^i$ (20 \% of $D_{\rm train}^{i-1}$). Then, a model $M'$ is trained on $D_A^i$ and then tested on $D_B^i$. The data in $D_B^i$ with lowest prediction errors (denoted as $D_{B, \rm unused}^i$) are then removed from the training set. Namely, $D_{\rm train}^{i} = D_{\rm train}^{i-1} - D_{B, \rm unused}^i$, and $D_{\rm unused}^{i}$ = $D_{\rm unused}^{i-1} + D_{B, \rm unused}^i$. The model $M^i$ trained on $D_{\rm train}^{i}$ is then used in the performance evaluation on the ID test set, the unused set $D_{\rm unused}^i$ and the OOD test set.
}


\subsection*{Active learning algorithm}
During the active learning process, the training set is initially constructed by randomly sampling 1~\% to 2~\% of the pool, and is grown with a batch size of 1~\% to 2~\% of the pool by selecting the materials with maximal prediction uncertainty. Three uncertainty measures are used to rank the materials. The first one is based on the uncertainty of the RF model and is calculated as the difference between the 95$^{th}$ and 5$^{th}$ percentile of the tree predictions in the forest. The second one is based on the uncertainty of the XGB model using an instance-based uncertainty estimation for gradient-boosted regression trees developed in Ref.~\cite{Brophy2022}. The third one is based on the query by committee, where the uncertainty is taken as the difference between the RF and XGB predictions.

\section*{Data availability}
The data required and generated by our code are available on Zenodo at url (to be inserted upon acceptance of the paper).

\section*{Code availability}
The code used in this work is available on GitHub at url (to be inserted upon acceptance of the paper).

\bibliography{lib_new} 

\section*{Acknowledgments}

The computations were made on the resources provided by the Calcul Quebec, Westgrid, and Compute Ontario consortia in the Digital Research Alliance of Canada (alliancecan.ca), and the Acceleration Consortium (acceleration.utoronto.ca) at the University of Toronto. We acknowledge funding provided by Natural Resources Canada’s Office of Energy Research and Development (OERD).  \copyright His Majesty the King in Right of Canada, as represented by the Minister of Natural Resources, 2023.

\section*{Author contributions}
K.L. and J.H.-S. conceived and designed the project. K.L. implemented the pruning algorithm. D.P. implemented the active learning algorithms. K.L. performed the ML training, analyzed the results, and drafted the manuscript. J.H.-S. supervised the project. All authors discussed the results, reviewed and edited the manuscript, and contributed to the manuscript preparation.

\section*{Competing interests}
The authors declare no competing interests.

\end{document}


\title{Supplemental information: \textit{On the redundancy in large material datasets: efficient and robust learning with less data}}

\author{Kangming Li\orcidlink{0000-0003-4471-8527}}
\affiliation{Department of Materials Science and Engineering, University of Toronto, 27 King’s College Cir, Toronto, ON, Canada.}

\author{Daniel Persaud}
\affiliation{Department of Materials Science and Engineering, University of Toronto, 27 King’s College Cir, Toronto, ON, Canada.}

\author{Kamal Choudhary\orcidlink{0000-0001-9737-8074}}
\affiliation{Material Measurement Laboratory, National Institute of Standards and Technology, 100 Bureau Dr, Gaithersburg, MD, USA.}
\affiliation{Theiss Research, La Jolla, CA 92037, USA.}

\author{Brian DeCost\orcidlink{0000-0002-3459-5888}}
\affiliation{Material Measurement Laboratory, National Institute of Standards and Technology, 100 Bureau Dr, Gaithersburg, MD, USA.}

\author{Michael Greenwood}
\affiliation{Canmet MATERIALS, Natural Resources Canada, 183 Longwood Road south, Hamilton, ON, Canada.}

\author{Jason Hattrick-Simpers\orcidlink{0000-0003-2937-3188}}
\email[Correspondence: ]{jason.hattrick.simpers@utoronto.ca}
\affiliation{Department of Materials Science and Engineering, University of Toronto, 27 King’s College Cir, Toronto, ON, Canada.}

\maketitle

In this document, we provide the supplementary information for: the model performance on in-distribution (ID) test data, the unused data in the pool and the out-of-distribution (OOD) test data in Sec.~\ref{sec:performance}; the label distribution of the pruned data sets in Sec.\ref{sec:label dist}; the transferability of pruned material sets in Sec.~\ref{sec:transferability}; the performance of the uncertainty-based active learning algorithms in Sec.~\ref{sec:AL}; the statistical overlaps between the older databases and OOD data in Sec.~\ref{sec:umap}.

\tableofcontents

\section{Model performance on ID, unused, and OOD test data~\label{sec:performance}}

In this section, we present the supplementary figures of the model performance, namely the root mean square errors (RMSE), and the coefficients of determination ($R^2$), for the formation energy and band gap predictions of the JARVIS, MP, and OQMD datasets. The model performance on the in-distribution (ID) test set, the unused data in the pool, and the out-of-distribution (OOD) test set is presented in Sec.~\ref{sec:ID}, Sec.~\ref{sec:unused}, and Sec.~\ref{sec:OOD}, respectively.

\subsection{Performance on ID test set\label{sec:ID}}
\begin{figure*}[htbp]
\centering
\includegraphics[width=0.48\linewidth]{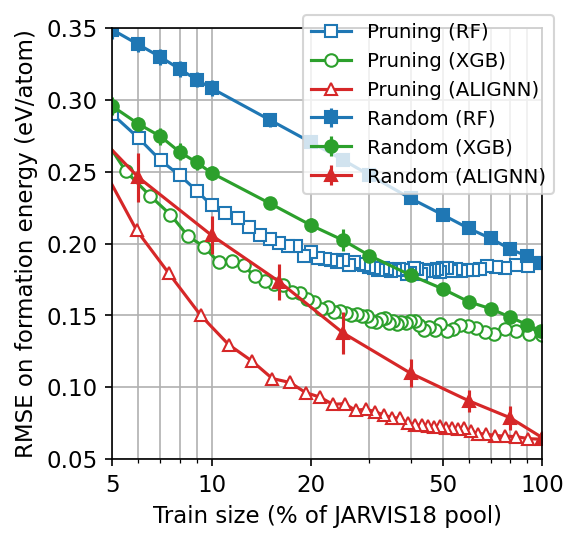}
\includegraphics[width=0.48\linewidth]{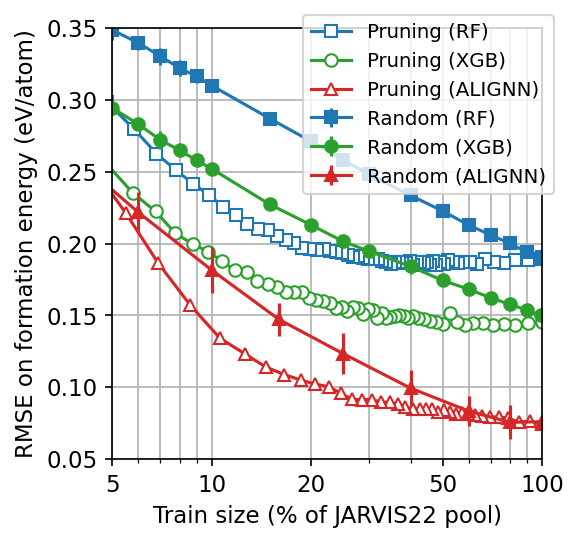}
\includegraphics[width=0.48\linewidth]{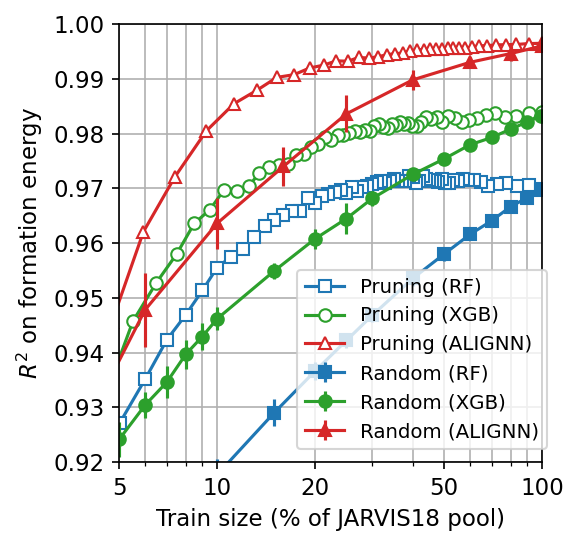}
\includegraphics[width=0.48\linewidth]{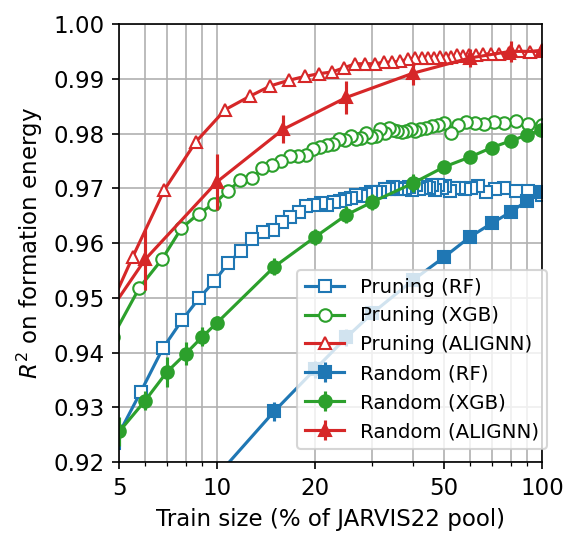}
\caption{\label{fig:ID performance e_form jarvis}RMSE (1st row) and $R^2$ (2nd row) on the ID test sets for the JARVIS18 (1st column) and JARVIS22 (2nd column) formation energy prediction. 
}
\end{figure*}

\begin{figure*}[htbp]
\centering
\includegraphics[width=0.48\linewidth]{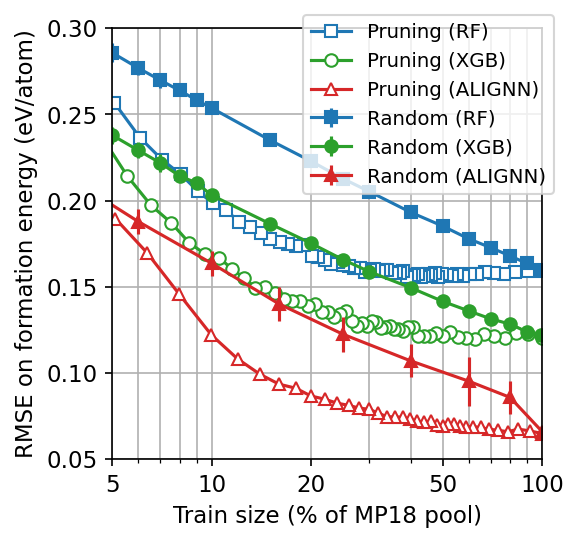}
\includegraphics[width=0.48\linewidth]{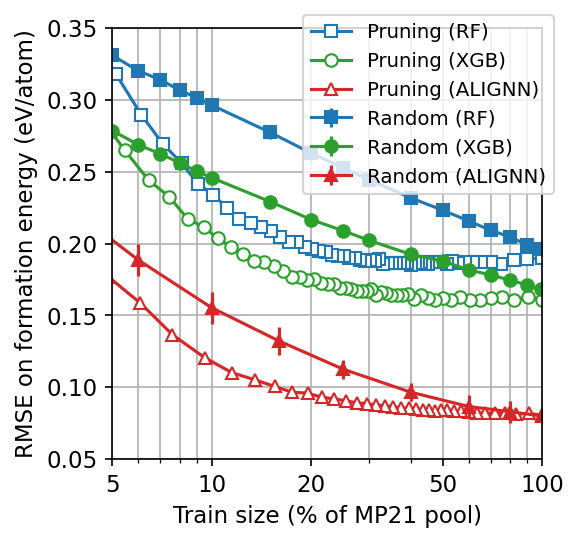}
\includegraphics[width=0.48\linewidth]{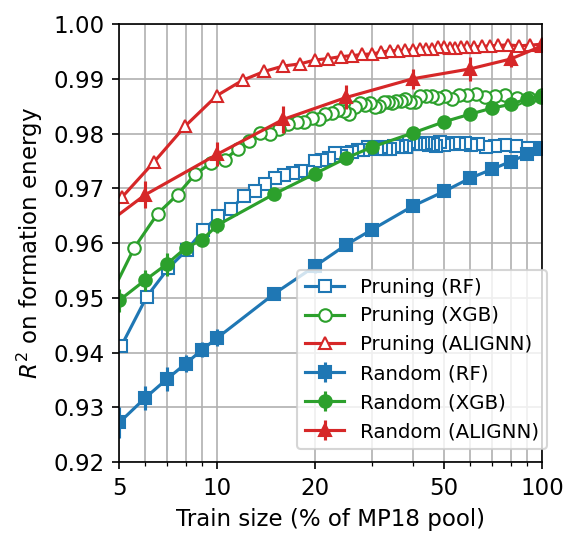}
\includegraphics[width=0.48\linewidth]{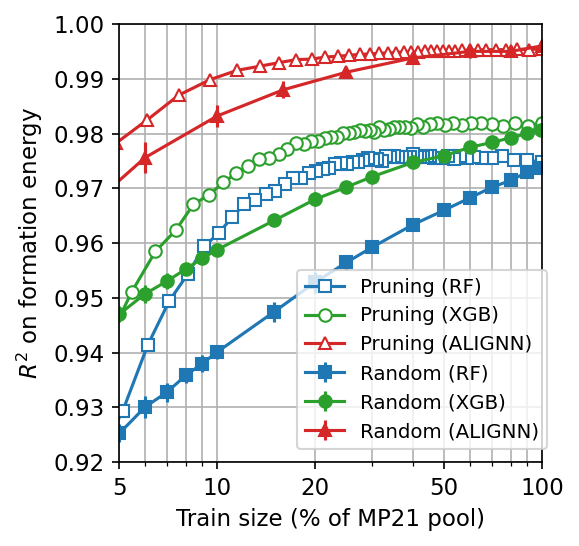}
\caption{\label{fig:ID performance e_form mp}RMSE (1st row) and $R^2$ (2nd row) on the ID test sets for the MP18 (1st column) and MP21 (2nd column) formation energy prediction. }
\end{figure*}

\begin{figure*}[htbp]
\centering
\includegraphics[width=0.48\linewidth]{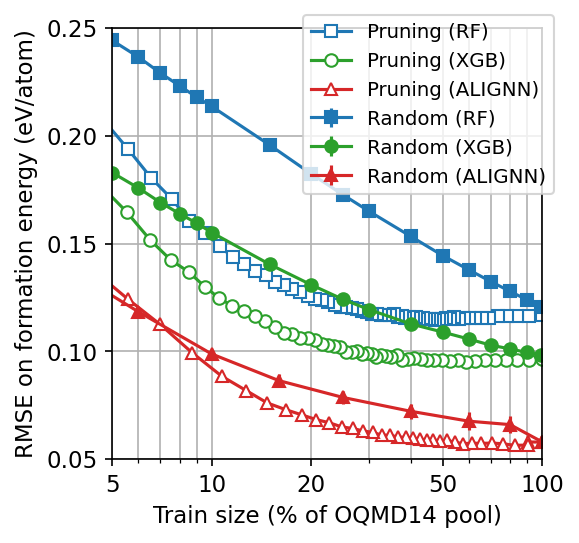}
\includegraphics[width=0.48\linewidth]{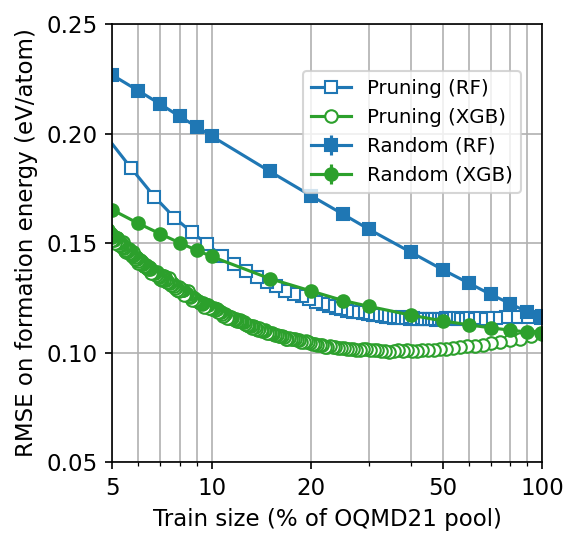}
\includegraphics[width=0.48\linewidth]{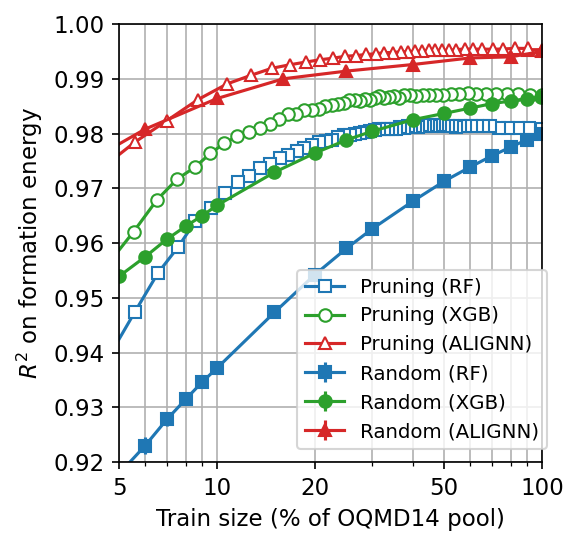}
\includegraphics[width=0.48\linewidth]{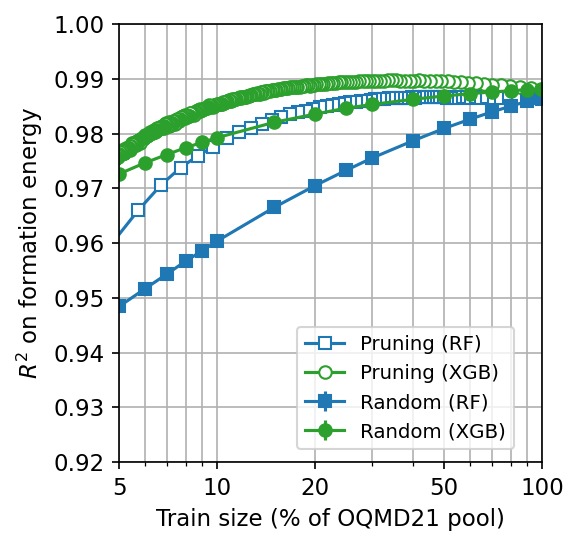}
\caption{\label{fig:ID performance e_form oqmd}RMSE (1st row) and $R^2$ (2nd row) on the ID test sets for the OQMD14 (1st column) and OQMD21 (2nd column) formation energy prediction. }
\end{figure*}

The ID performance (RMSE and $R^2$) of the formation energy models is shown in Fig.~\ref{fig:ID performance e_form jarvis} for the JARVIS18 and JARVIS22 datasets, Fig.~\ref{fig:ID performance e_form mp} for the MP18 and MP21 datasets, and Fig.~\ref{fig:ID performance e_form oqmd} for the OQMD14 and OQMD21 datasets.

\begin{figure*}[htbp]
\centering
\includegraphics[width=0.48\linewidth]{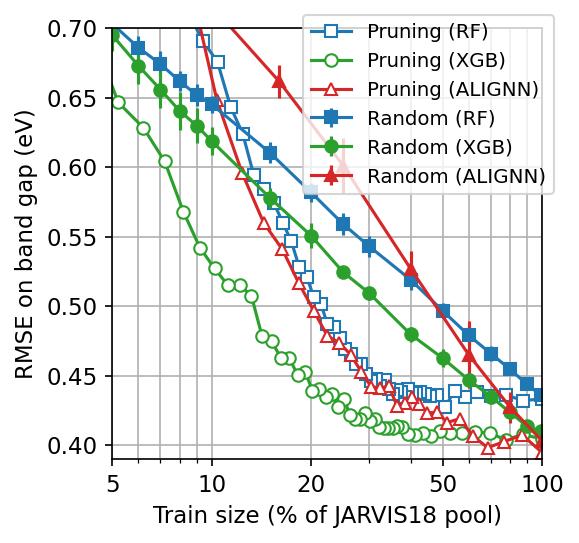}
\includegraphics[width=0.48\linewidth]{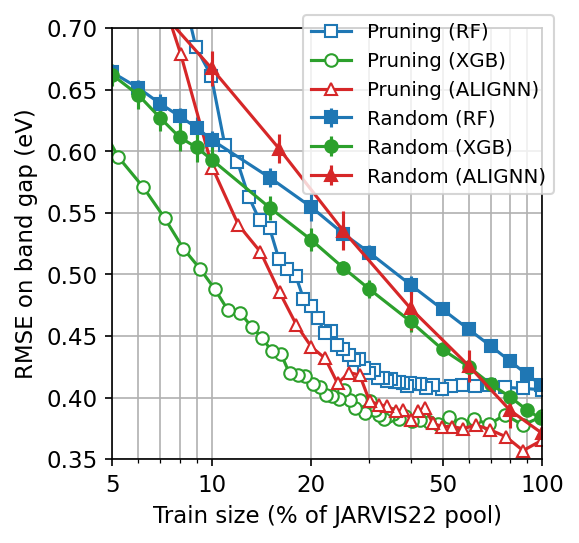}
\includegraphics[width=0.48\linewidth]{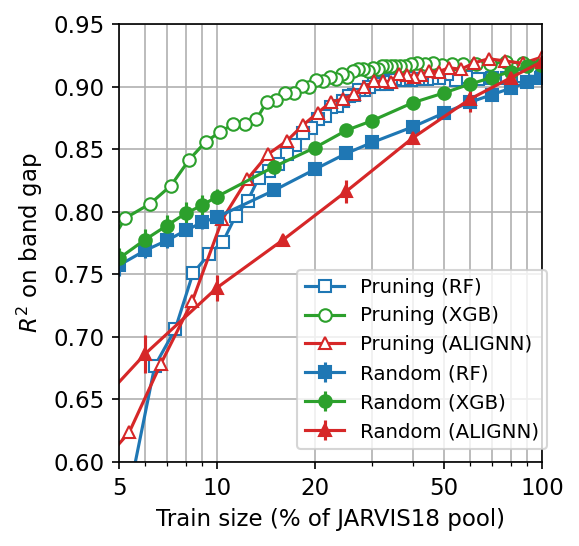}
\includegraphics[width=0.48\linewidth]{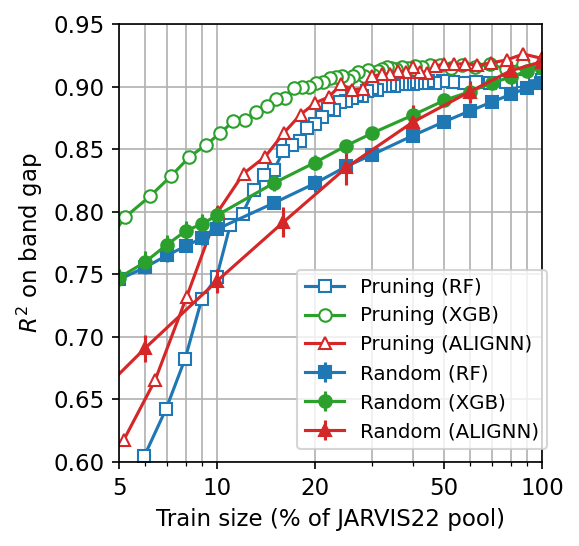}
\caption{\label{fig:ID performance bandgap jarvis}RMSE (1st row) and $R^2$ (2nd row) on the ID test sets for the JARVIS18 (1st column) and JARVIS22 (2nd column) band gap prediction. 
}
\end{figure*}

\begin{figure*}[htbp]
\centering
\includegraphics[width=0.48\linewidth]{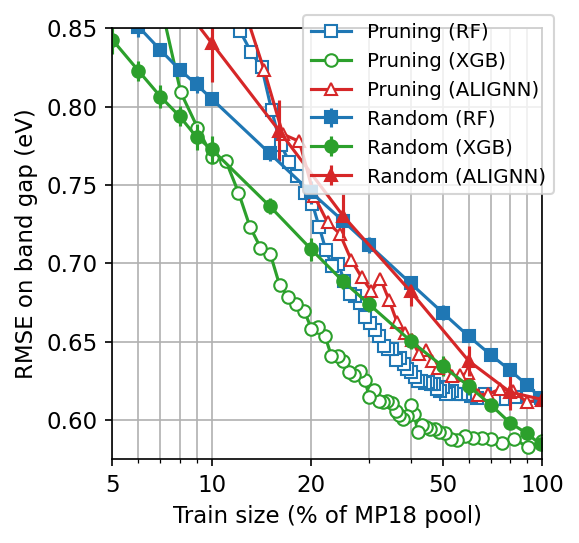}
\includegraphics[width=0.48\linewidth]{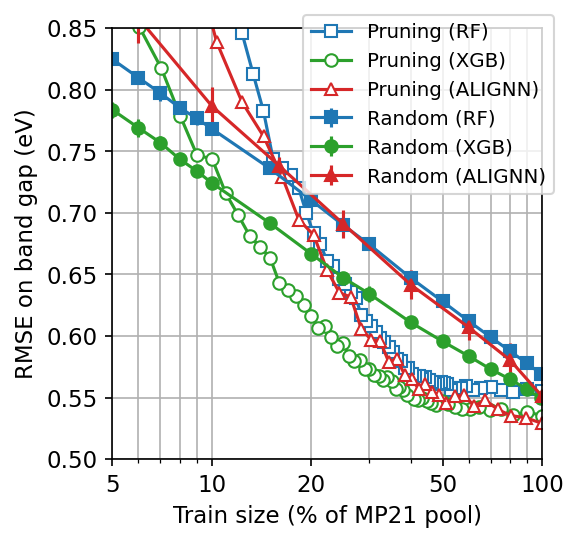}
\includegraphics[width=0.48\linewidth]{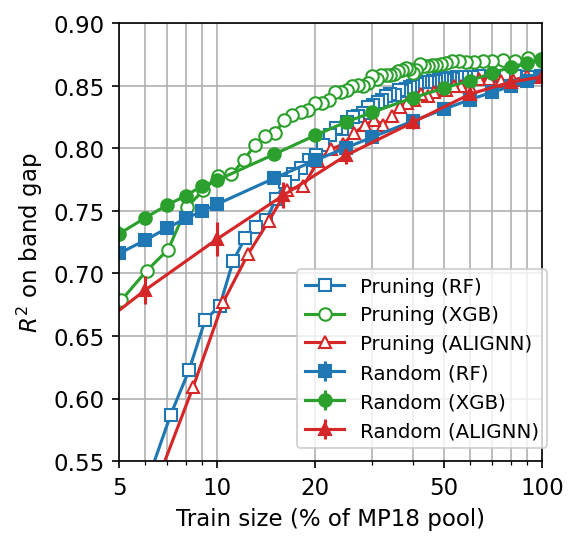}
\includegraphics[width=0.48\linewidth]{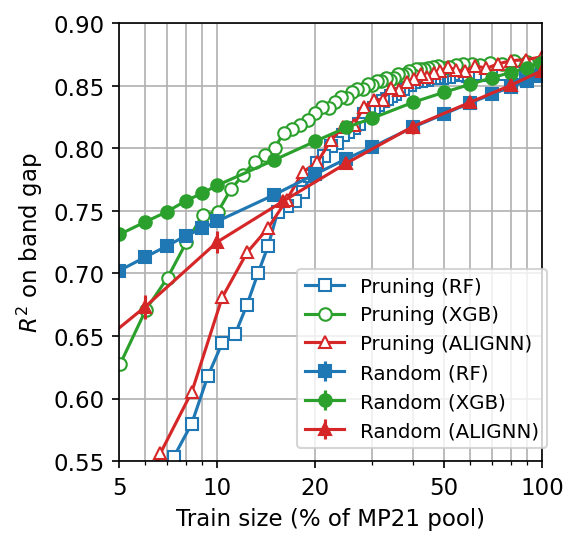}
\caption{\label{fig:ID performance bandgap mp}RMSE (1st row) and $R^2$ (2nd row) on the ID test sets for the MP18 (1st column) and MP21 (2nd column) band gap prediction. }
\end{figure*}

\begin{figure*}[htbp]
\centering
\includegraphics[width=0.48\linewidth]{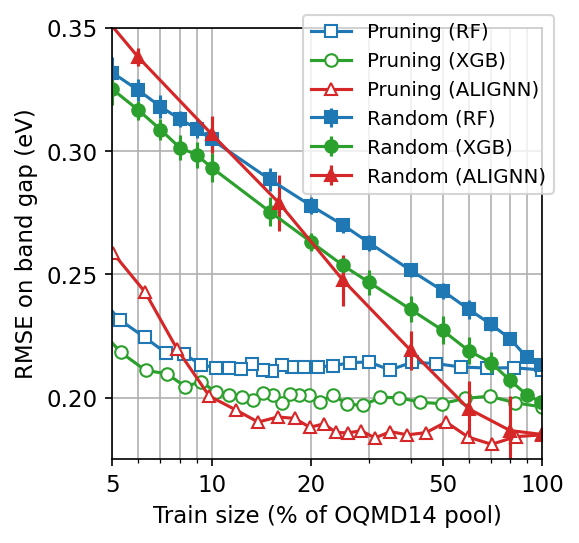}
\includegraphics[width=0.48\linewidth]{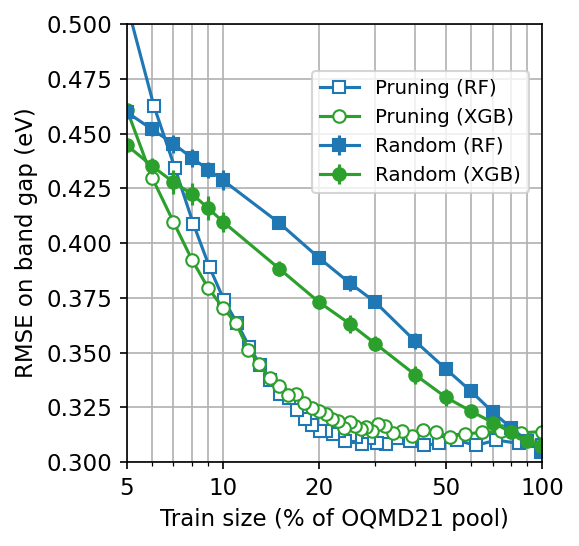}
\includegraphics[width=0.48\linewidth]{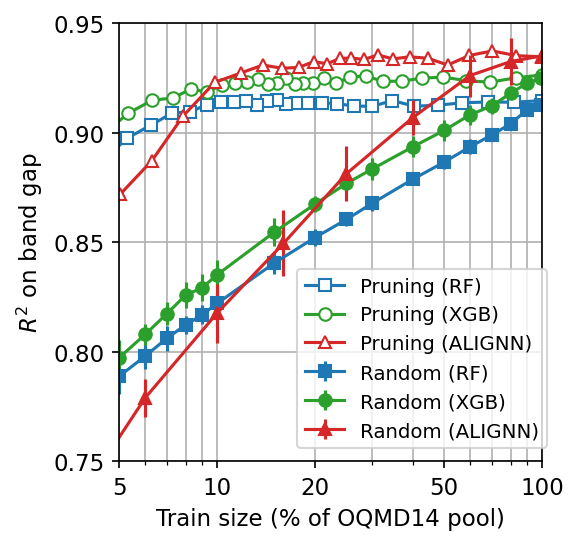}
\includegraphics[width=0.48\linewidth]{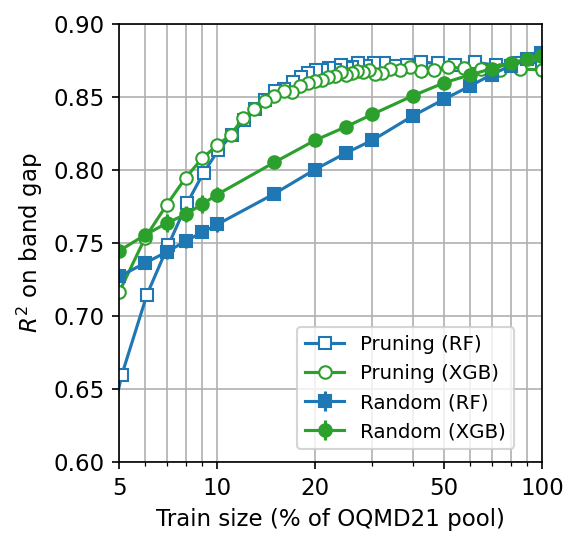}
\caption{\label{fig:ID performance bandgap oqmd}RMSE (1st row) and $R^2$ (2nd row) on the ID test sets for the OQMD14 (1st column) and OQMD21 (2nd column) band gap prediction. }
\end{figure*}

The ID performance (RMSE and $R^2$) of the band gap models is shown in Fig.~\ref{fig:ID performance bandgap jarvis} for the JARVIS18 and JARVIS22 datasets, Fig.~\ref{fig:ID performance bandgap mp} for the MP18 and MP21 datasets, and Fig.~\ref{fig:ID performance bandgap oqmd} for the OQMD14 and OQMD21 datasets.

\FloatBarrier
\subsection{Performance on unused data\label{sec:unused}}
\begin{figure*}[htbp]
\centering
\includegraphics[width=0.48\linewidth]{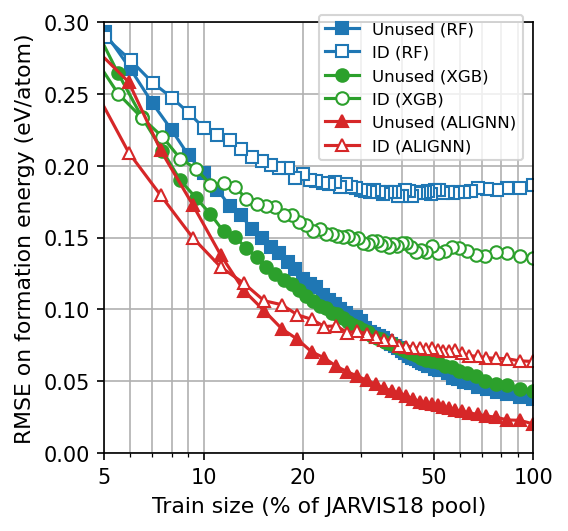}
\includegraphics[width=0.48\linewidth]{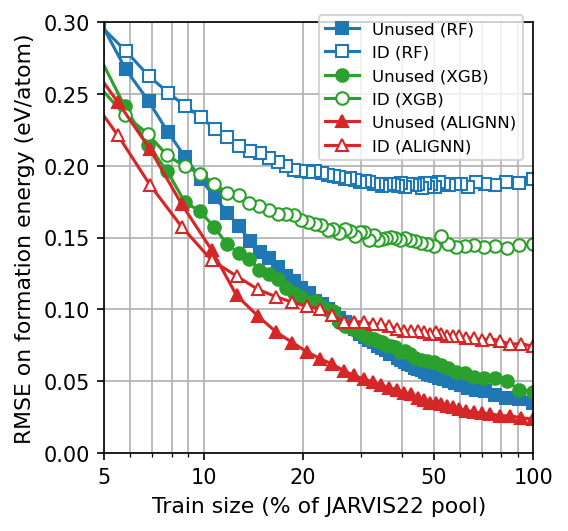}
\includegraphics[width=0.48\linewidth]{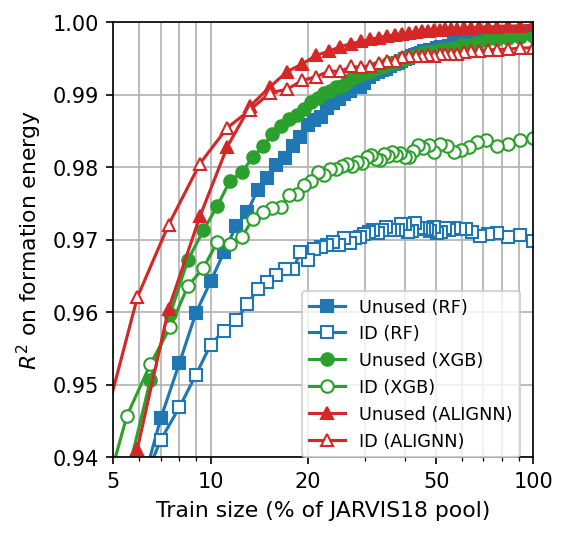}
\includegraphics[width=0.48\linewidth]{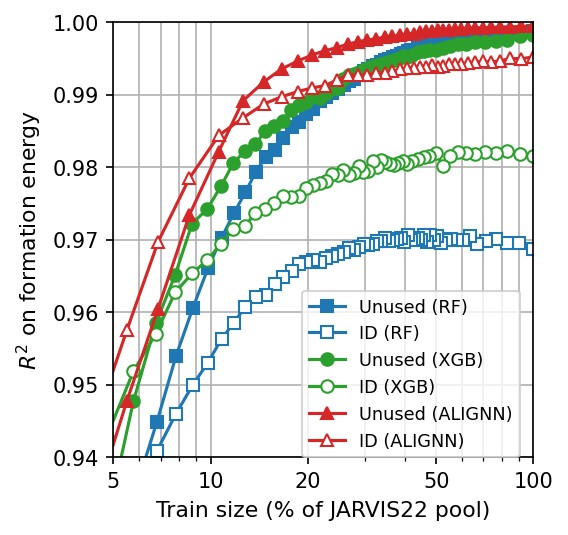}
\caption{\label{fig:val performance e_form jarvis}RMSE (1st row) and $R^2$ (2nd row) on the unused data for the JARVIS18 (1st column) and JARVIS22 (2nd column) formation energy prediction. 
}
\end{figure*}

\begin{figure*}[htbp]
\centering
\includegraphics[width=0.48\linewidth]{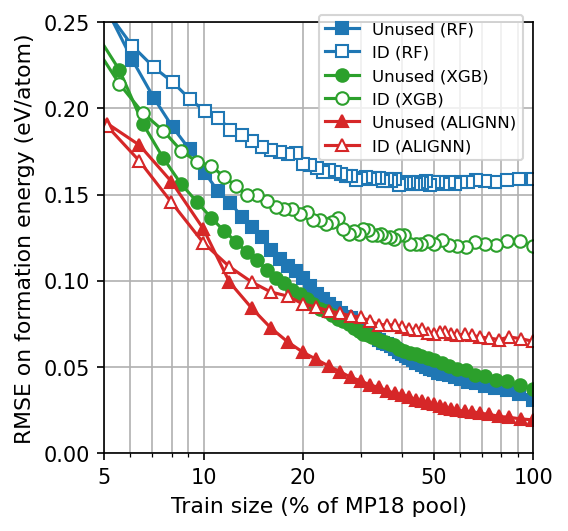}
\includegraphics[width=0.48\linewidth]{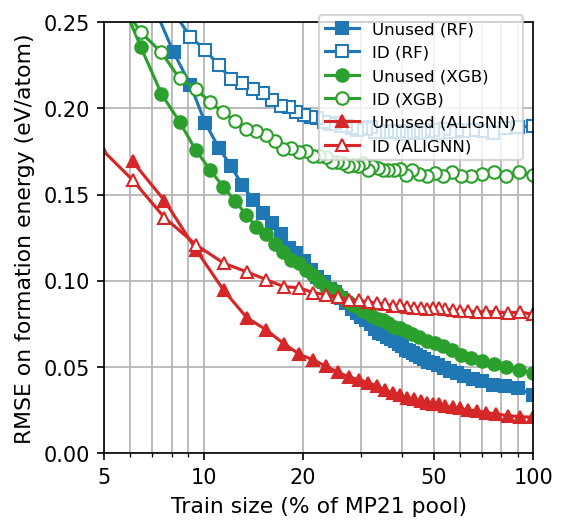}
\includegraphics[width=0.48\linewidth]{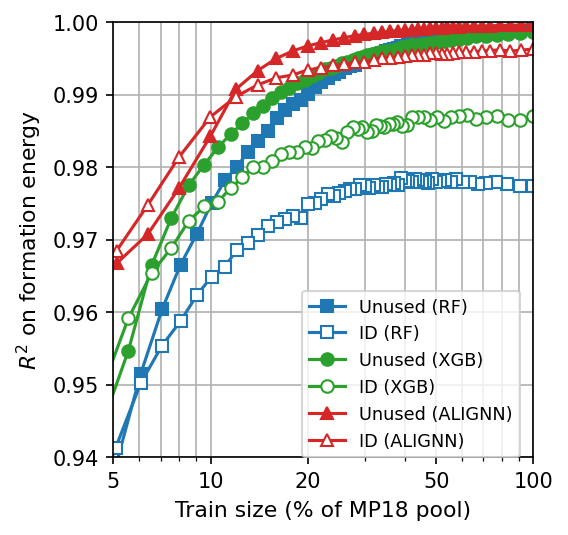}
\includegraphics[width=0.48\linewidth]{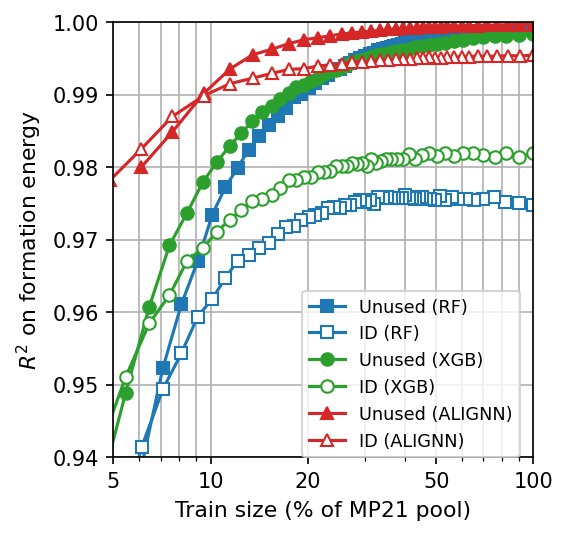}
\caption{\label{fig:val performance e_form mp}RMSE (1st row) and $R^2$ (2nd row) on the unused data for the MP18 (1st column) and MP21 (2nd column) formation energy prediction. }
\end{figure*}

\begin{figure*}[htbp]
\centering
\includegraphics[width=0.48\linewidth]{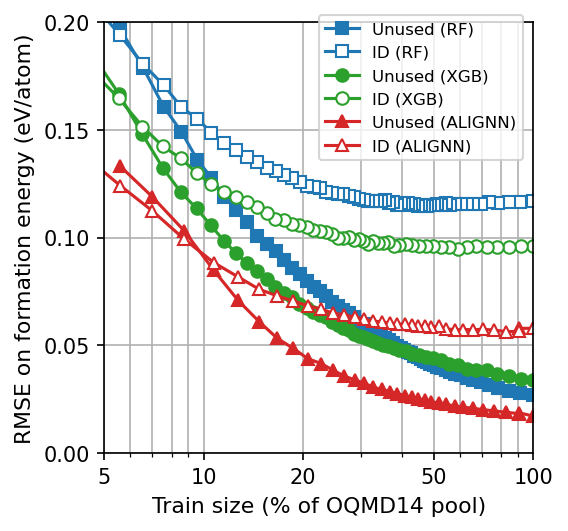}
\includegraphics[width=0.48\linewidth]{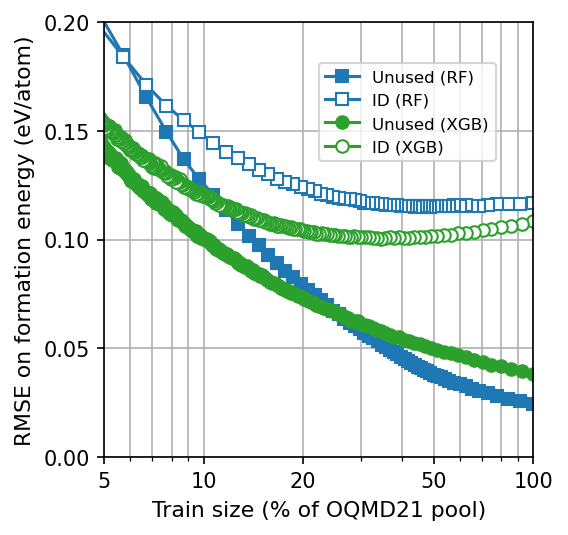}
\includegraphics[width=0.48\linewidth]{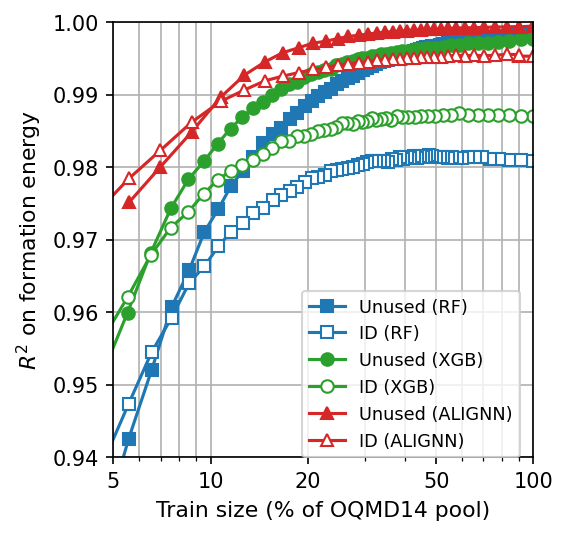}
\includegraphics[width=0.48\linewidth]{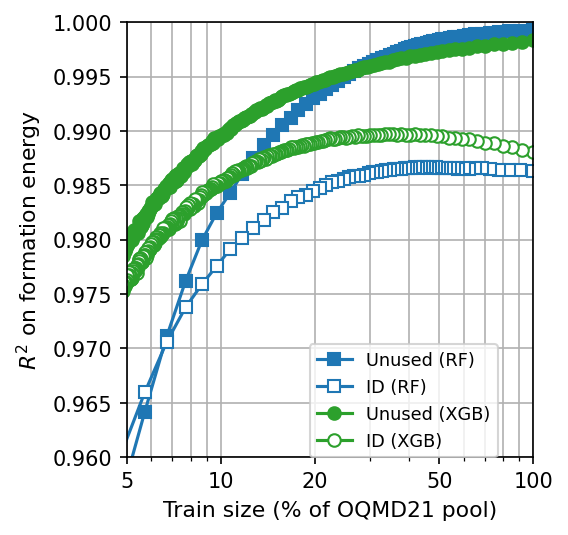}
\caption{\label{fig:val performance e_form oqmd}RMSE (1st row) and $R^2$ (2nd row) on the unused data for the OQMD14 (1st column) and OQMD21 (2nd column) formation energy prediction. }
\end{figure*}

The performance (RMSE and $R^2$) of the formation energy models on the unused data is shown in Fig.~\ref{fig:val performance e_form jarvis} for the JARVIS18 and JARVIS22 datasets, Fig.~\ref{fig:val performance e_form mp} for the MP18 and MP21 datasets, and Fig.~\ref{fig:val performance e_form oqmd} for the OQMD14 and OQMD21 datasets.

\begin{figure*}[htbp]
\centering
\includegraphics[width=0.48\linewidth]{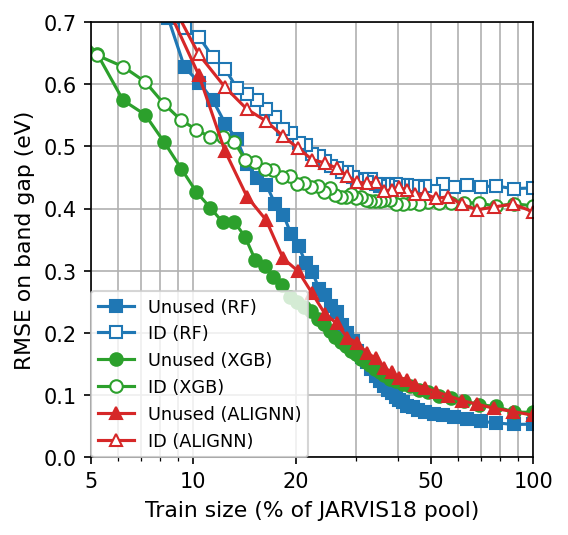}
\includegraphics[width=0.48\linewidth]{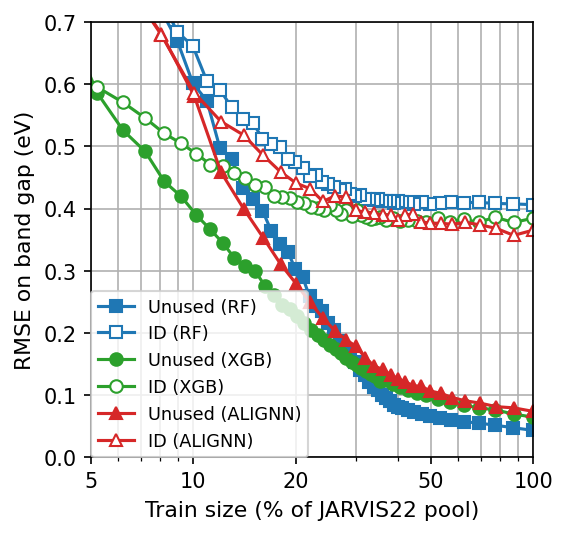}
\includegraphics[width=0.48\linewidth]{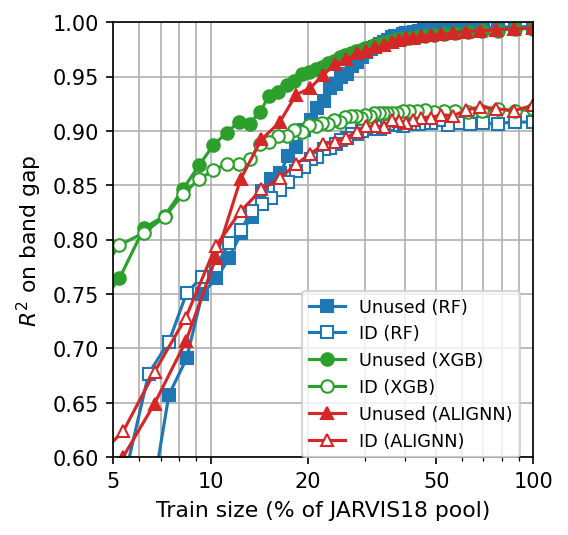}
\includegraphics[width=0.48\linewidth]{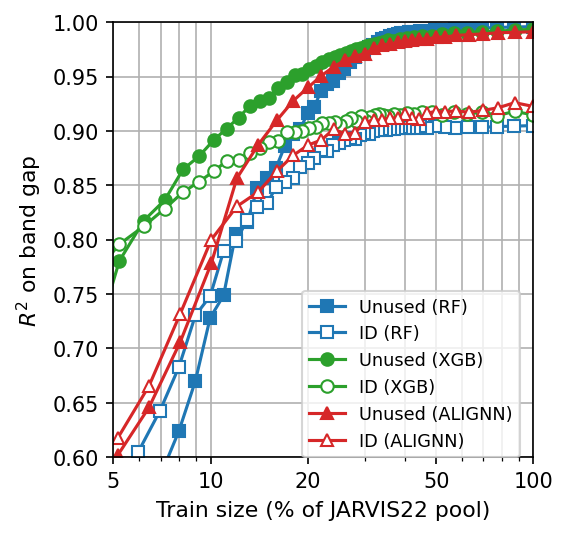}
\caption{\label{fig:val performance bandgap jarvis}RMSE (1st row) and $R^2$ (2nd row) on the unused data for the JARVIS18 (1st column) and JARVIS22 (2nd column) band gap prediction. 
}
\end{figure*}

\begin{figure*}[htbp]
\centering
\includegraphics[width=0.48\linewidth]{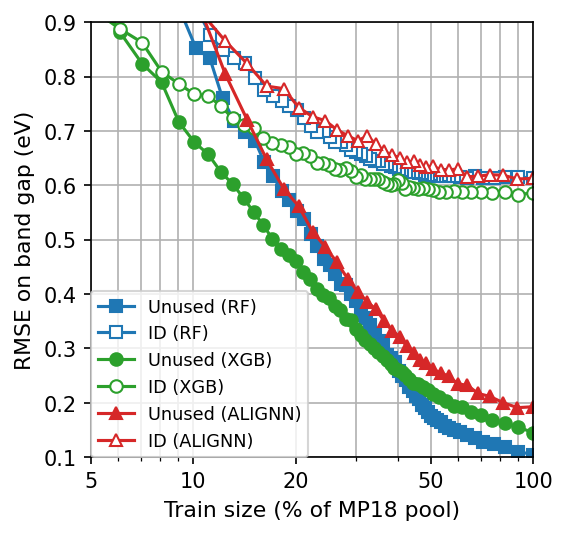}
\includegraphics[width=0.48\linewidth]{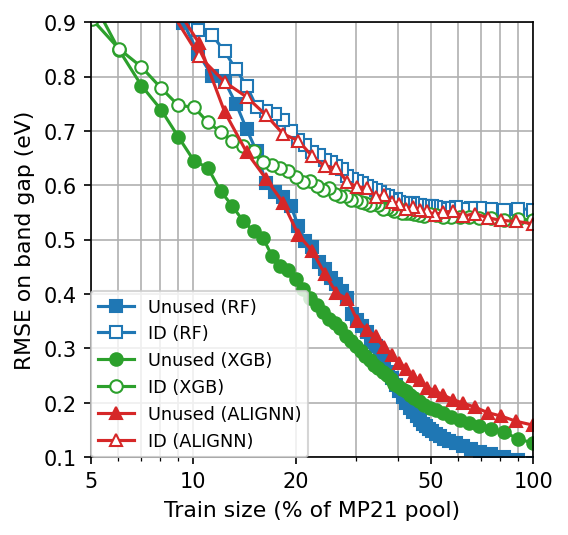}
\includegraphics[width=0.48\linewidth]{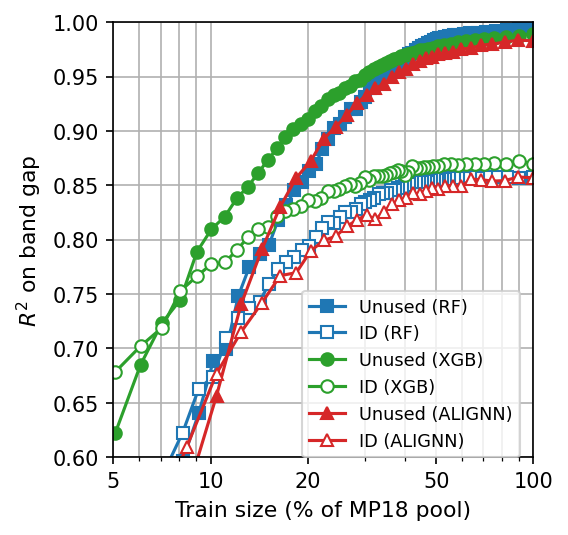}
\includegraphics[width=0.48\linewidth]{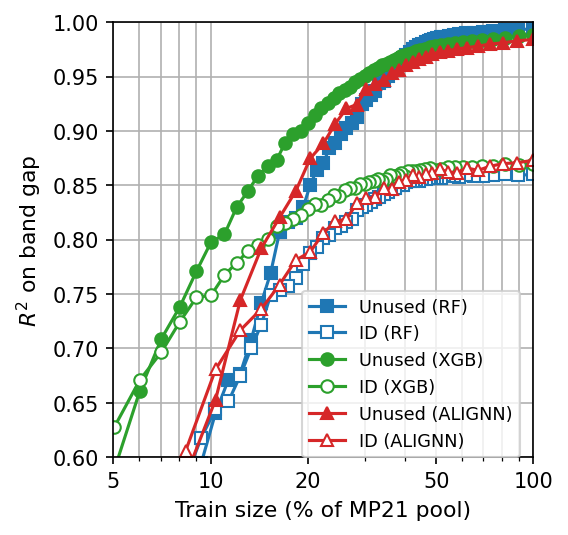}
\caption{\label{fig:val performance bandgap mp}RMSE (1st row) and $R^2$ (2nd row) on the unused data for the MP18 (1st column) and MP21 (2nd column) band gap prediction. }
\end{figure*}

\begin{figure*}[htbp]
\centering
\includegraphics[width=0.48\linewidth]{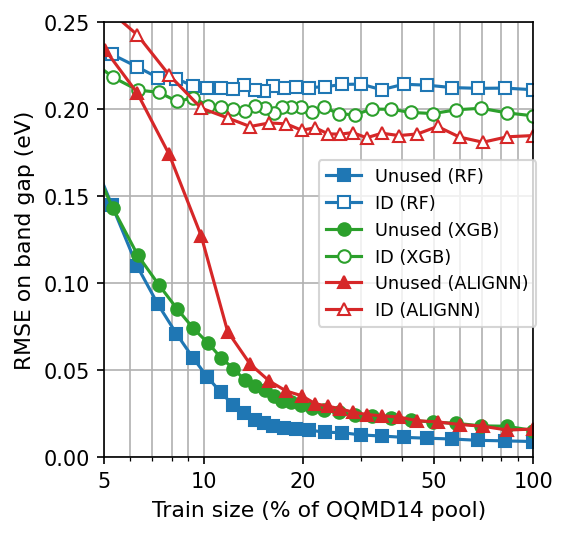}
\includegraphics[width=0.48\linewidth]{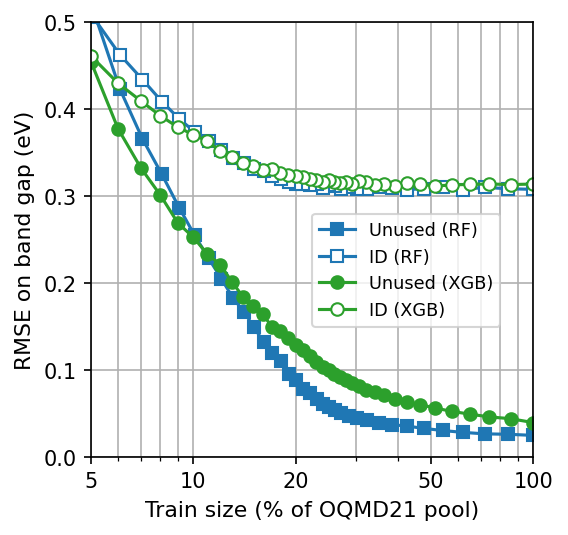}
\includegraphics[width=0.48\linewidth]{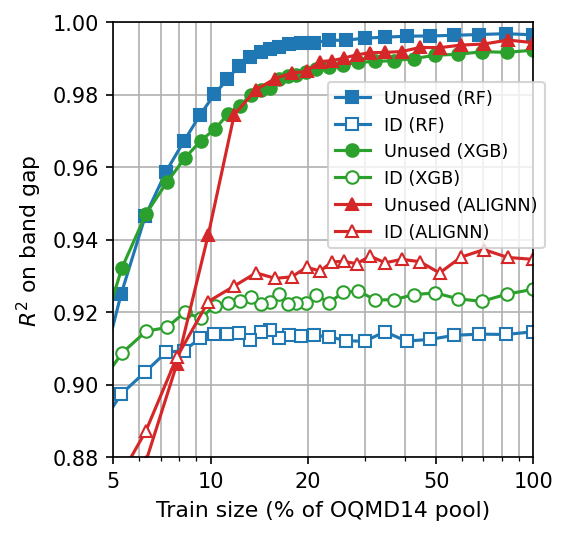}
\includegraphics[width=0.48\linewidth]{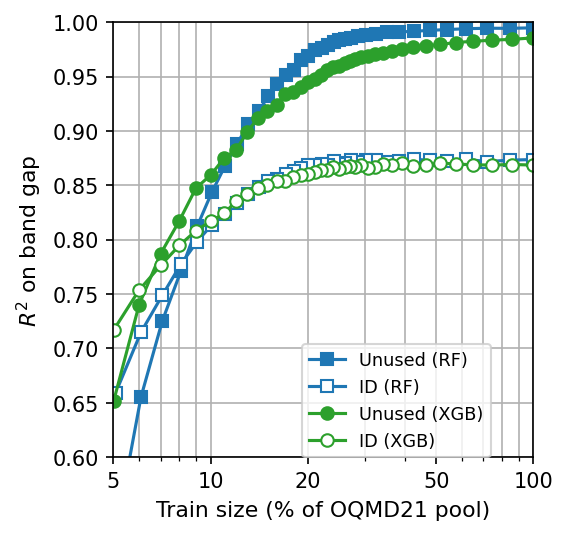}
\caption{\label{fig:val performance bandgap oqmd}RMSE (1st row) and $R^2$ (2nd row) on the unused data for the OQMD14 (1st column) and OQMD21 (2nd column) band gap prediction. }
\end{figure*}

The performance (RMSE and $R^2$) of the band gap models on the unused data is shown in Fig.~\ref{fig:val performance bandgap jarvis} for the JARVIS18 and JARVIS22 datasets, Fig.~\ref{fig:val performance bandgap mp} for the MP18 and MP21 datasets, and Fig.~\ref{fig:val performance bandgap oqmd} for the OQMD14 and OQMD21 datasets.

\FloatBarrier

\subsection{Performance on OOD test set\label{sec:OOD}}

\begin{figure*}[!htbp]
\centering
\includegraphics[width=0.4\linewidth]{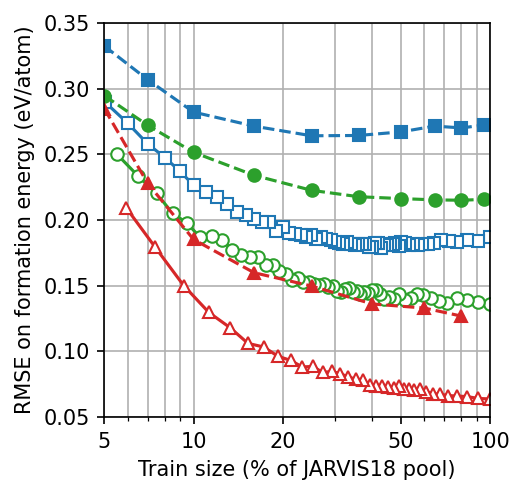}
\includegraphics[width=0.4\linewidth]{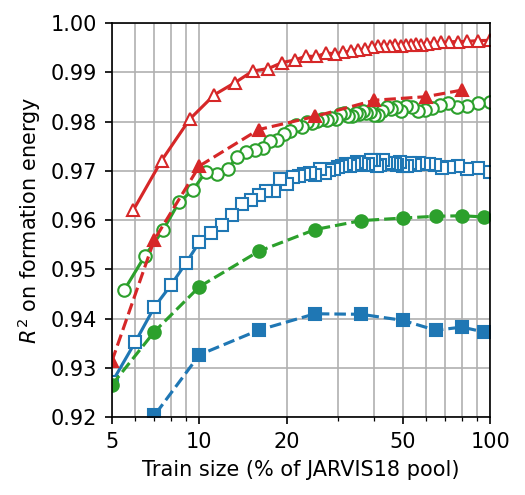}
\includegraphics[width=0.4\linewidth]{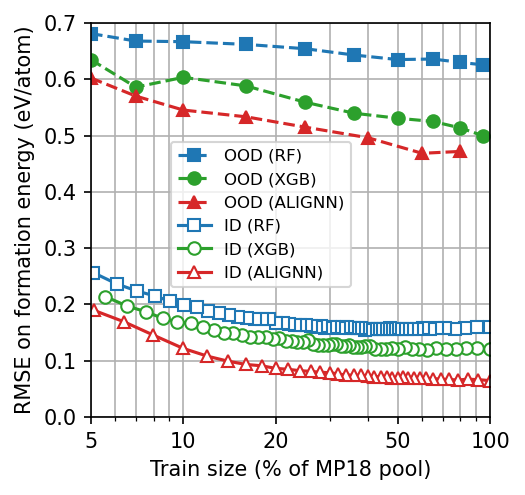}
\includegraphics[width=0.4\linewidth]{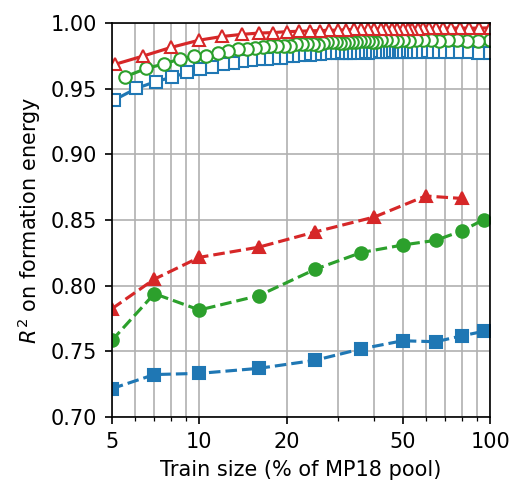}
\includegraphics[width=0.4\linewidth]{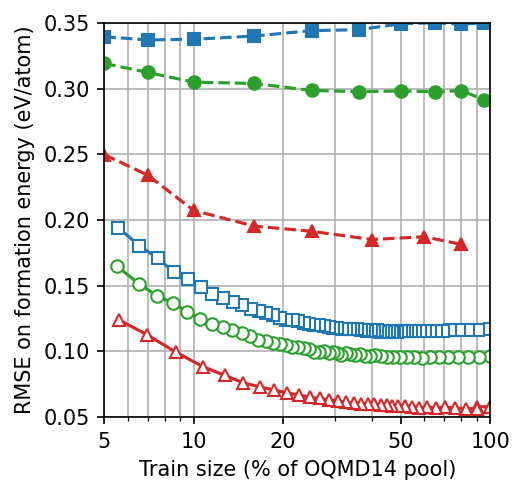}
\includegraphics[width=0.4\linewidth]{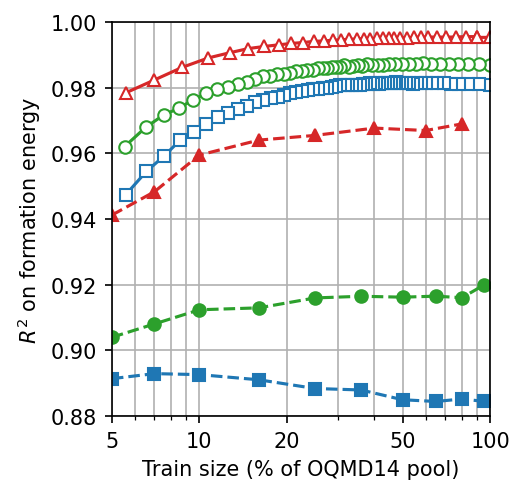}
\caption{\label{fig:ood performance e_form}
OOD performance of the JARVIS, MP, and OQMD formation energy predictions.
}
\end{figure*}

\begin{figure*}[!htbp]
\centering
\includegraphics[width=0.4\linewidth]{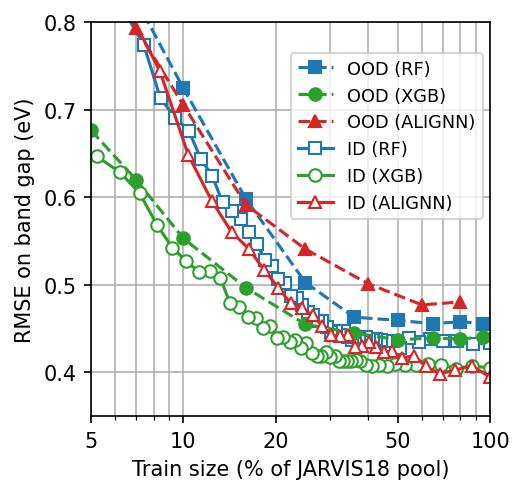}
\includegraphics[width=0.4\linewidth]{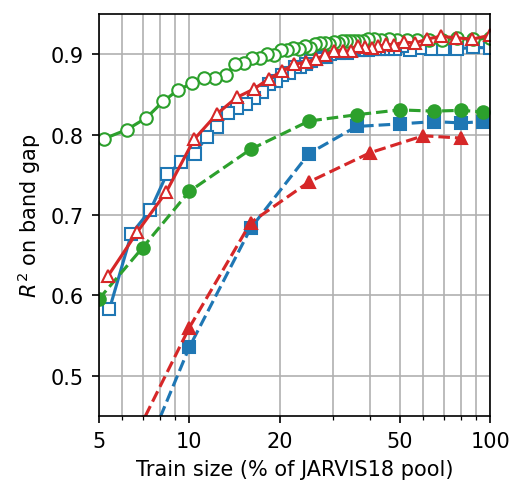}
\includegraphics[width=0.4\linewidth]{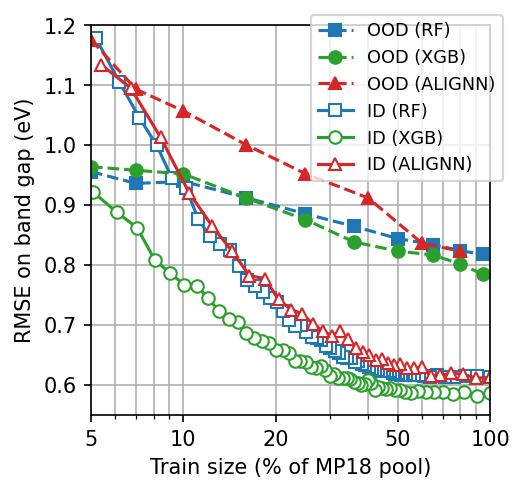}
\includegraphics[width=0.4\linewidth]{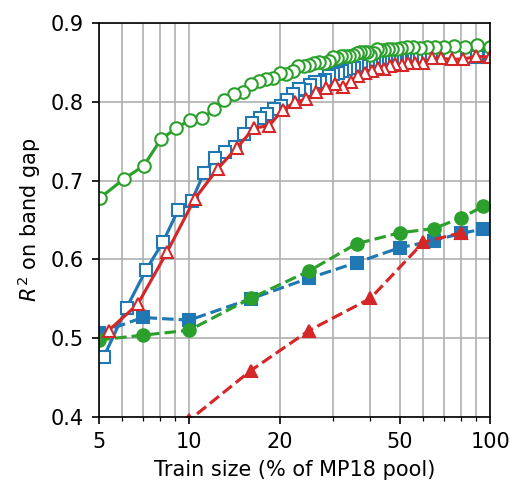}
\includegraphics[width=0.4\linewidth]{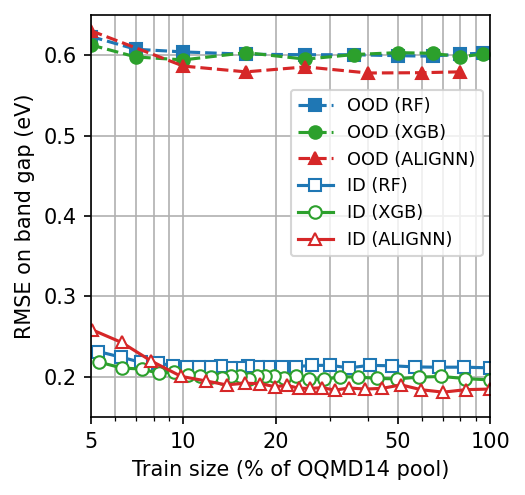}
\includegraphics[width=0.4\linewidth]{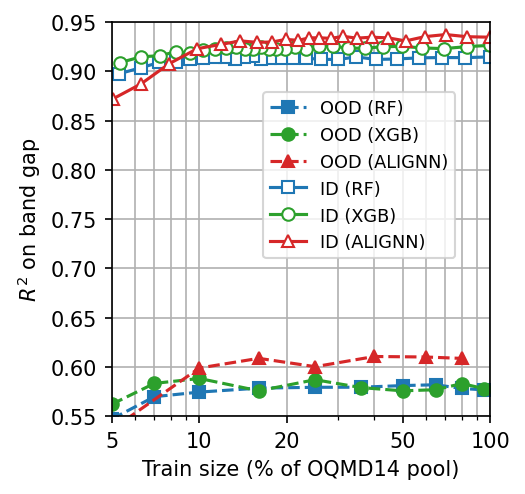}
\caption{\label{fig:ood performance bandgap}
OOD performance of the JARVIS, MP, and OQMD band gap predictions.
}
\end{figure*}

To quantify the degree of the performance degradation due to the distribution shift, we train the models using the entire pool of the older datasets (JARVIS18, MP18, or OQMD14), and test their performance on the hold-out ID test sets of the older datasets, and on the OOD test sets in the newer datsets (JARVIS22, MP21, or OQMD21). The OOD performance (RMSE and $R^2$) of the formation energy models is shown in Fig.~\ref{fig:ood performance e_form} for the JARVIS, MP, and OQMD datasets. The OOD performance (RMSE and $R^2$) of the band gap models is shown in Fig.~\ref{fig:ood performance bandgap} for the JARVIS, MP, and OQMD datasets. The ratios of the ID RMSE to the OOD RMSE for the models trained on 100 \% of the pool are given in Table~\ref{tab:ood vs id}.

\begin{table}[htbp]
\caption{The ratio of the RMSE on the OOD test set to that on the ID test set using the full models (namely, trained on the entire pool). }
\label{tab:ood vs id}
  \begin{ruledtabular}
\begin{tabular}{ccccc}
Database                & Property         & RF  & XGB & ALIGNN \\ \hline
\multirow{2}{*}{JARVIS} & formation energy & 1.5 & 1.6 & 2.0    \\
                        & band gap         & 1.1 & 1.1 & 1.2    \\
\multirow{2}{*}{MP}     & formation energy & 4.0 & 4.2 & 7.3    \\
                        & band gap         & 1.3 & 1.3 & 1.3    \\
\multirow{2}{*}{OQMD}   & formation energy & 3.1 & 3.1 & 3.2    \\
                        & band gap         & 2.9 & 3.1 & 3.2   
\end{tabular}
  \end{ruledtabular}
\end{table}

\FloatBarrier
\section{Label distribution of the pruned data sets\label{sec:label dist}}

\begin{figure*}[!htbp]
\centering
\includegraphics[width=0.4\linewidth]{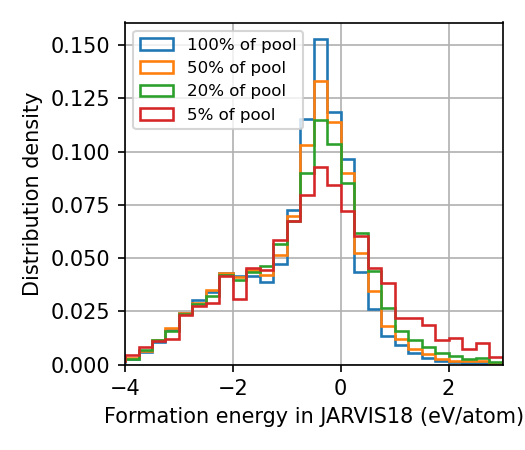}
\includegraphics[width=0.4\linewidth]{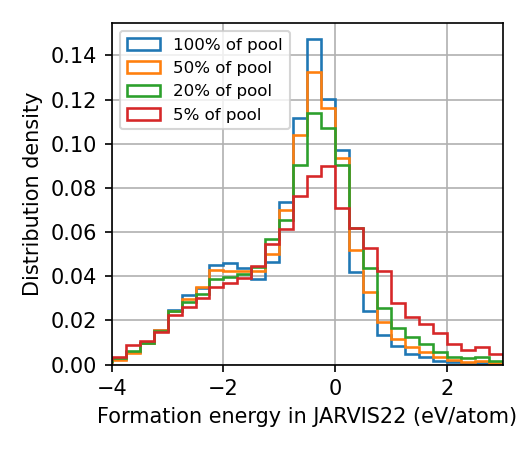}
\includegraphics[width=0.4\linewidth]{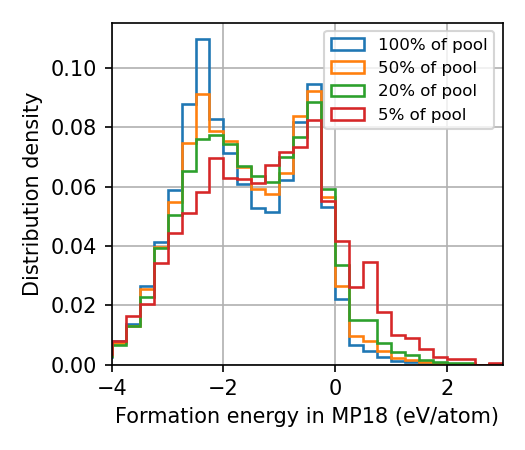}
\includegraphics[width=0.4\linewidth]{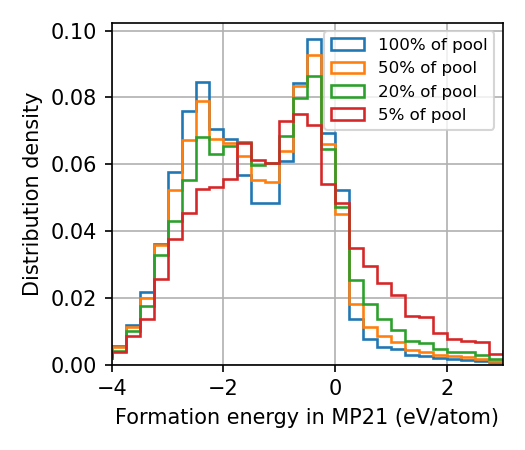}
\includegraphics[width=0.4\linewidth]{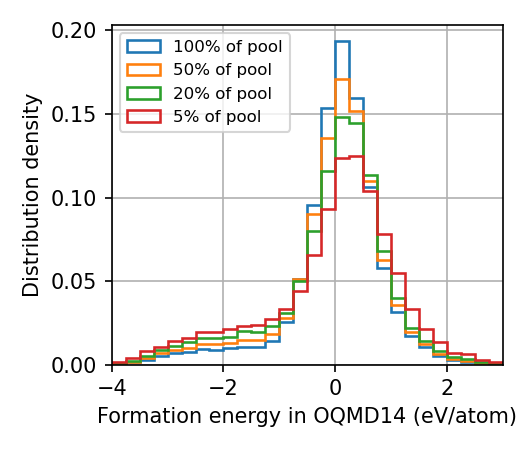}
\includegraphics[width=0.4\linewidth]{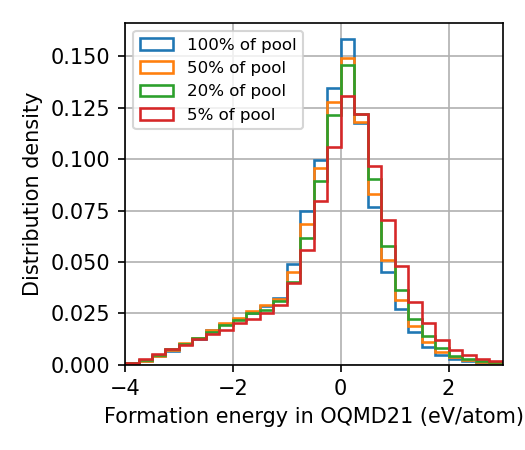}
\caption{\label{fig:label dist e_form}
Label distribution of the XGB-pruned formation energy data. In each figure, the distributions of the training sets accounting for 100 \% to 5 \% of the pool are shown. 
}
\end{figure*}

\begin{figure*}[!htbp]
\centering
\includegraphics[width=0.4\linewidth]{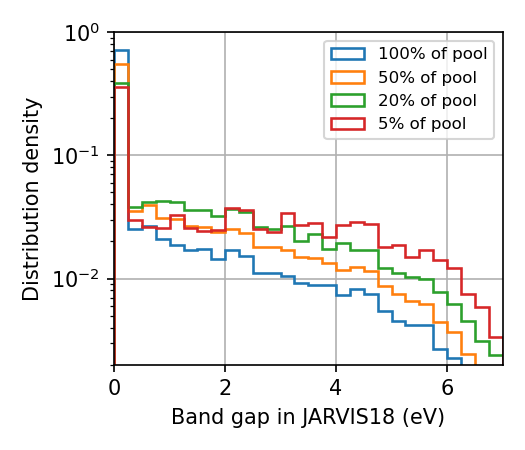}
\includegraphics[width=0.4\linewidth]{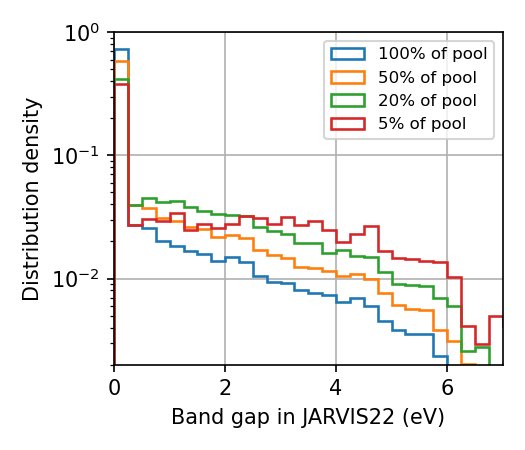}
\includegraphics[width=0.4\linewidth]{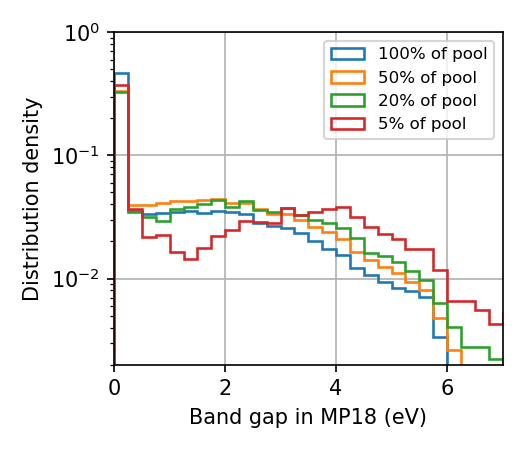}
\includegraphics[width=0.4\linewidth]{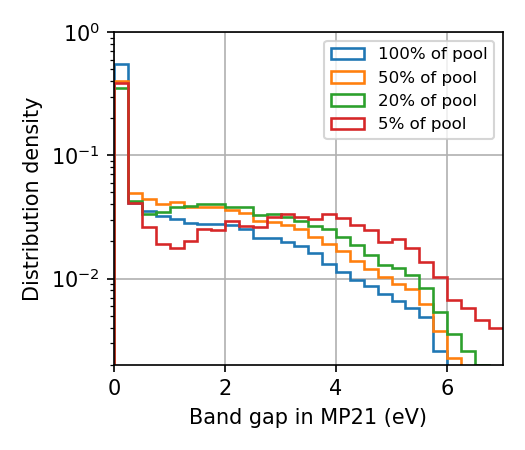}
\includegraphics[width=0.4\linewidth]{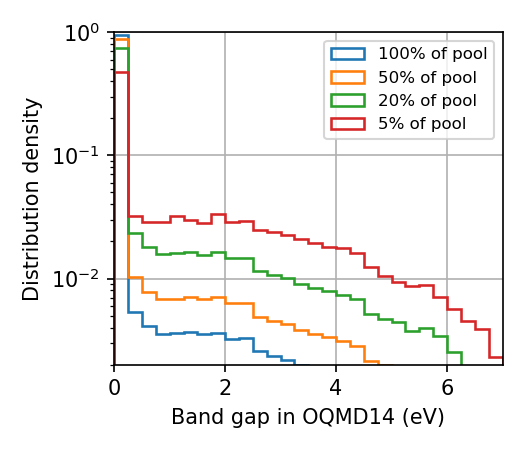}
\includegraphics[width=0.4\linewidth]{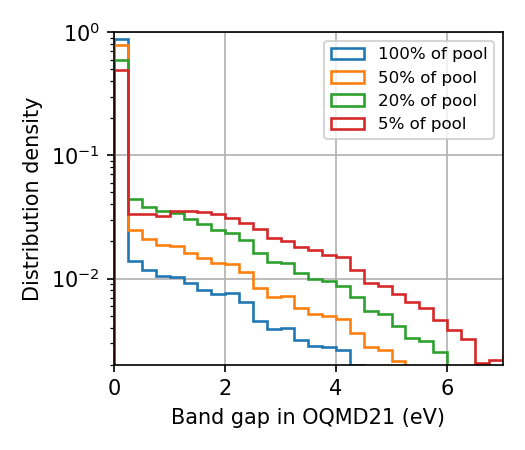}
\caption{\label{fig:label dist bandgap}
Label distribution of the XGB-pruned band gap data. In each figure, the distributions of the training sets accounting for 100 \% to 5 \% of the pool are shown. Please note that the distribution density is in the logarithmic scale.
}
\end{figure*}

The pruned data exhibits a distribution different from the original distribution of $S_0$. To demonstrate this point, we show the label distributions of the XGB-pruned formation energy data in Fig.~\ref{fig:label dist e_form} and band gap data in Fig.~\ref{fig:label dist bandgap}. Compared to the original distributions (100 \% of the pool), the distributions of the pruned data (50 \%, 20 \%, and 5 \% of the pool) are increasingly skewed towards less stable materials which are underrepresented in the original distribution. Similarly for the band gap data, a large portion of the materials have a band gap close to zero in the original distribution, whereas the distributions of the pruned data are skewed towards materials with larger band gaps. 

\FloatBarrier
\section{Transferability of pruned material sets\label{sec:transferability}}
\subsection{Transferability between ML models}

\begin{figure*}[!htbp]
\centering
\includegraphics[width=0.4\linewidth]{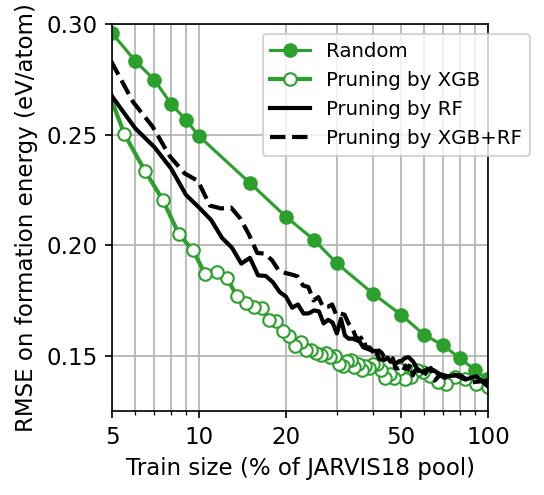}
\includegraphics[width=0.4\linewidth]{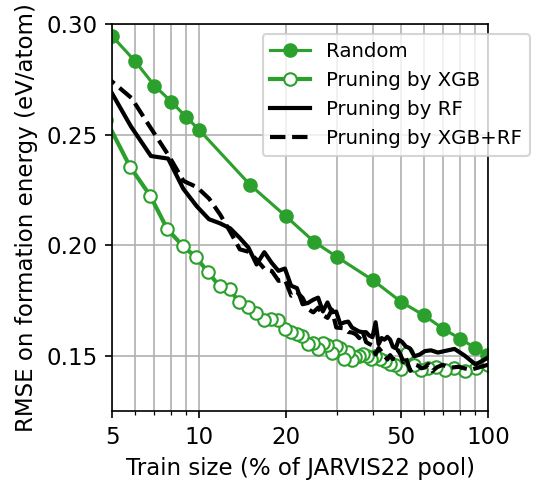}
\includegraphics[width=0.4\linewidth]{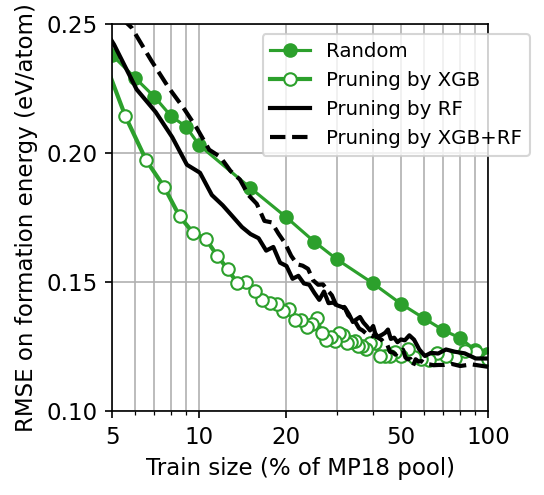}
\includegraphics[width=0.4\linewidth]{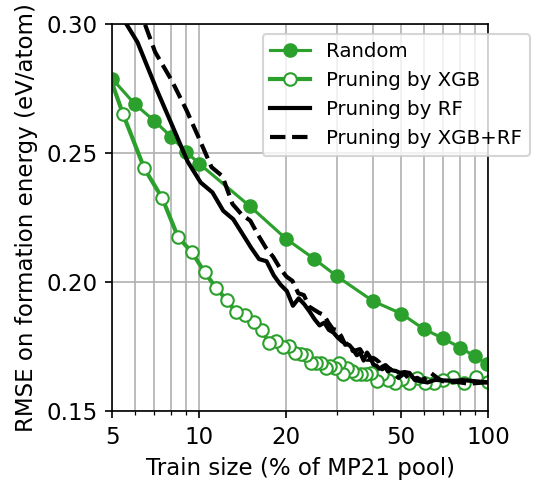}
\includegraphics[width=0.4\linewidth]{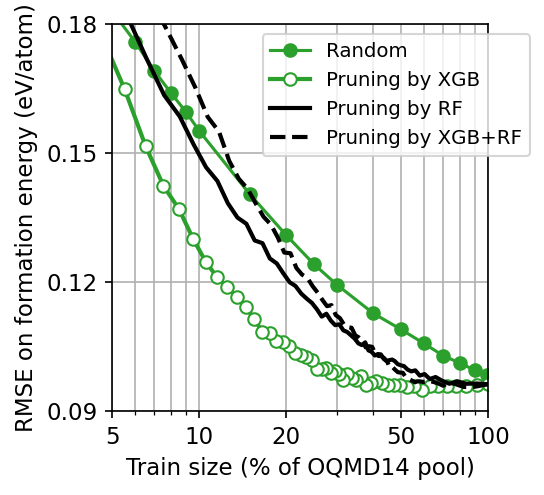}
\includegraphics[width=0.4\linewidth]{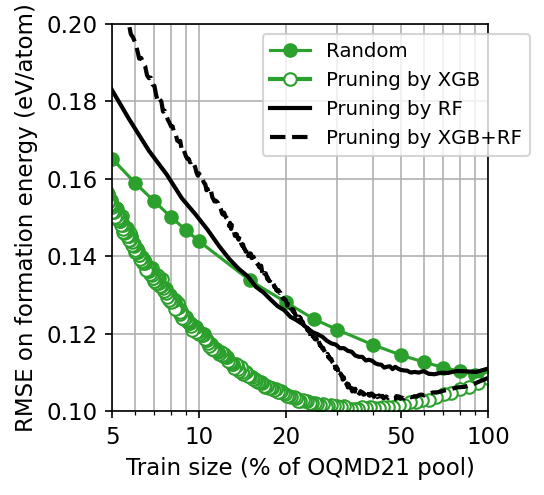}
\caption{\label{fig:trans_ml xgb e_form}
ID performance of the XGB models for JARVIS18, JARVIS22, MP18, MP21, OQMD14, and OQMD21 formation energy datasets. For each dataset, the RMSE scores obtained by training the XGB models on the randomly selected data, the data pruned by the XGB models, the data pruned by the RF models, and the data jointly pruned by the XGB and RF models are shown.
}
\end{figure*}

\begin{figure*}[!htbp]
\centering
\includegraphics[width=0.4\linewidth]{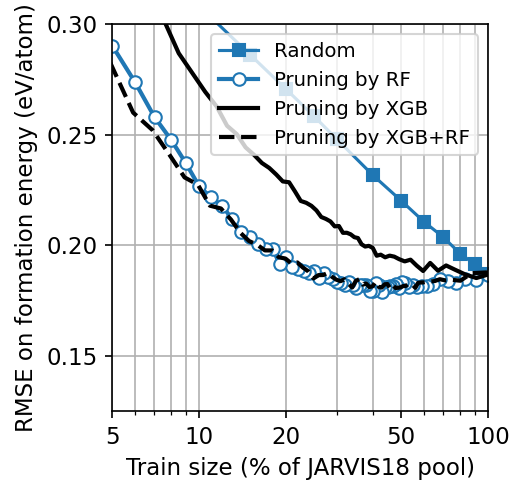}
\includegraphics[width=0.4\linewidth]{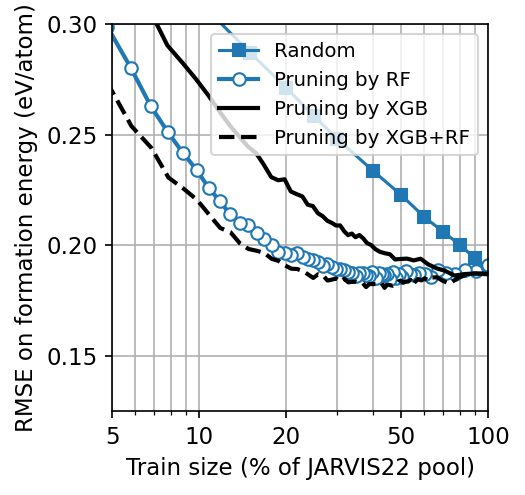}
\includegraphics[width=0.4\linewidth]{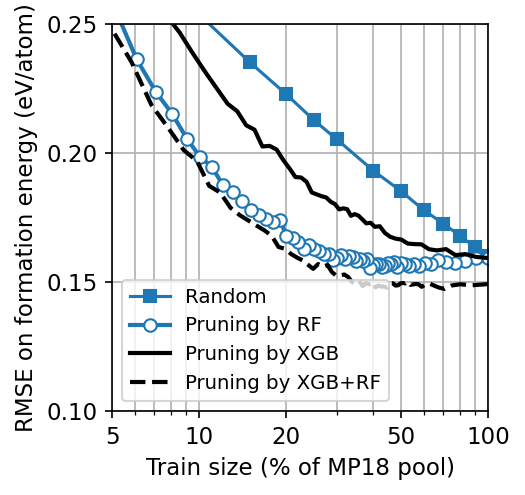}
\includegraphics[width=0.4\linewidth]{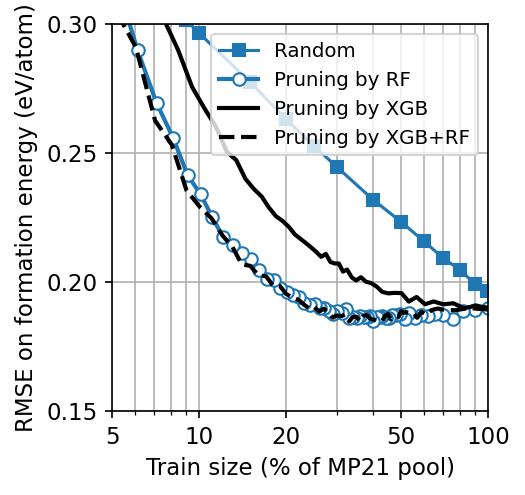}
\includegraphics[width=0.4\linewidth]{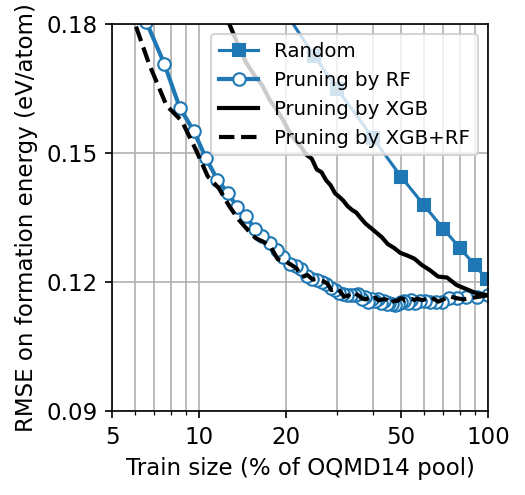}
\includegraphics[width=0.4\linewidth]{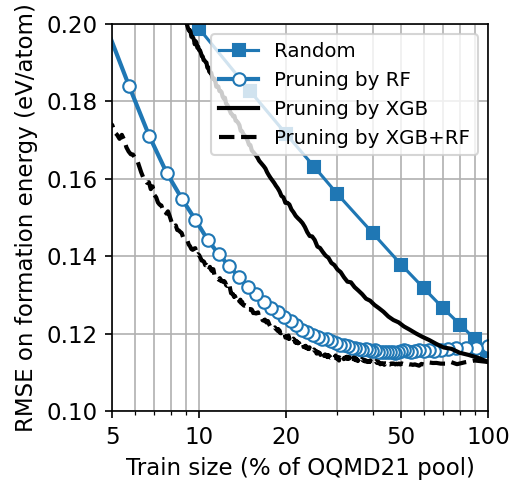}
\caption{\label{fig:trans_ml rf e_form}
ID performance of the RF models for JARVIS18, JARVIS22, MP18, MP21, OQMD14, and OQMD21 formation energy datasets. For each dataset, the RMSE scores obtained by training the RF models on the randomly selected data, the data pruned by the RF models, the data pruned by the XGB models, and the data jointly pruned by the XGB and RF models are shown.
}
\end{figure*}

\begin{figure*}[!htbp]
\centering
\includegraphics[width=0.4\linewidth]{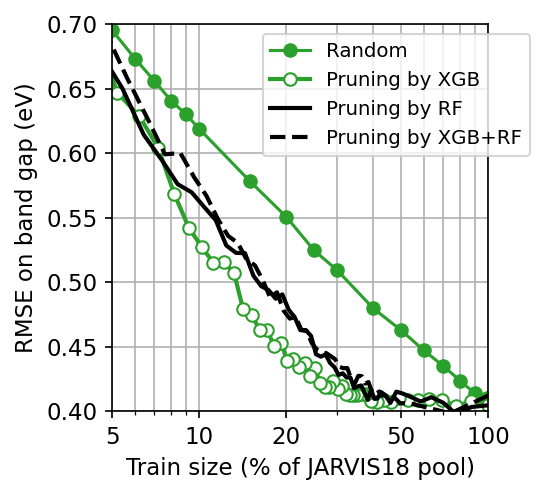}
\includegraphics[width=0.4\linewidth]{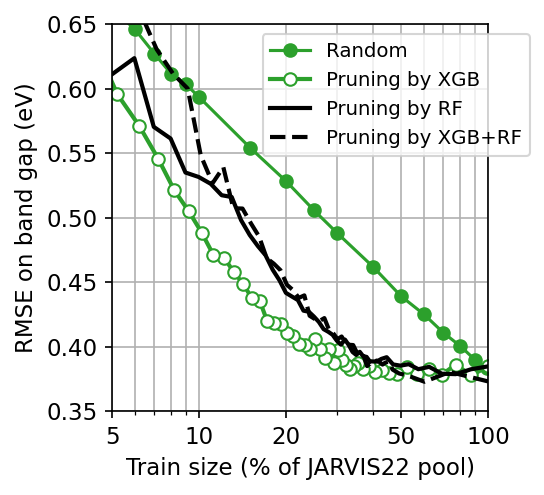}
\includegraphics[width=0.4\linewidth]{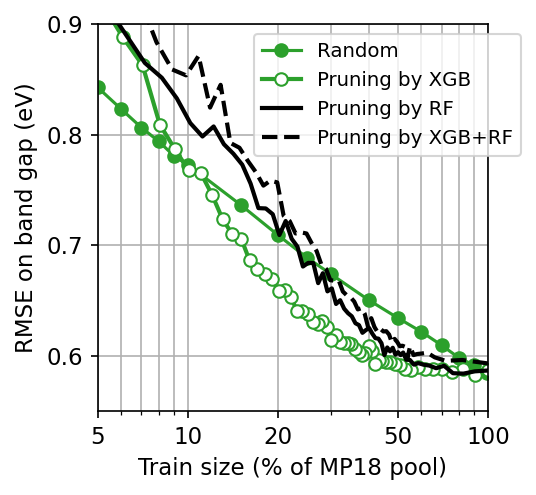}
\includegraphics[width=0.4\linewidth]{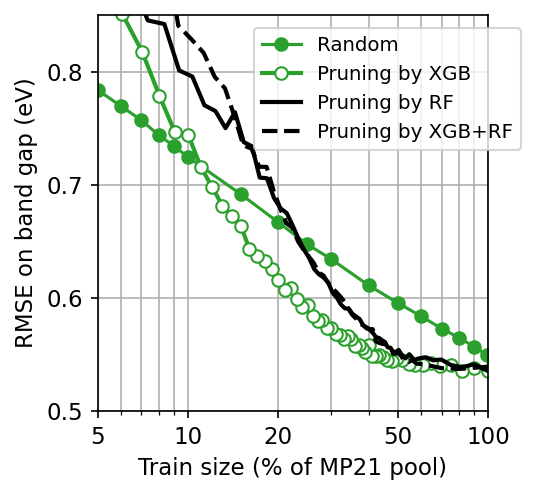}
\includegraphics[width=0.4\linewidth]{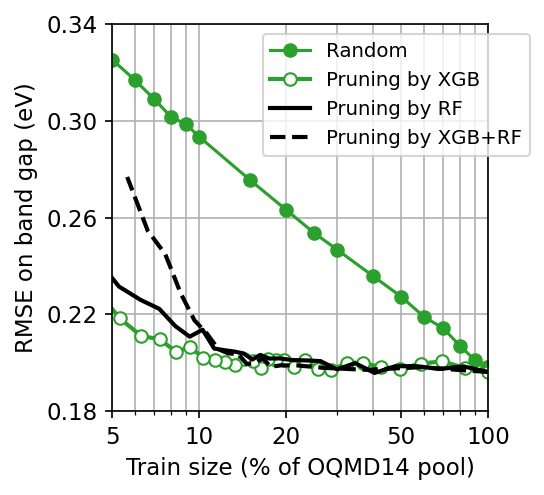}
\includegraphics[width=0.4\linewidth]{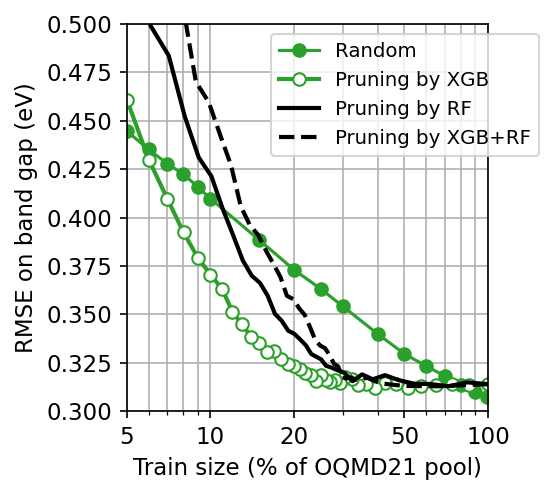}
\caption{\label{fig:trans_ml xgb bandgap}
ID performance of the XGB models for JARVIS18, JARVIS22, MP18, MP21, OQMD14, and OQMD21 band gap datasets. For each dataset, the RMSE scores obtained by training the XGB models on the randomly selected data, the data pruned by the XGB models, the data pruned by the RF models, and the data jointly pruned by the XGB and RF models are shown.
}
\end{figure*}

\begin{figure*}[!htbp]
\centering
\includegraphics[width=0.4\linewidth]{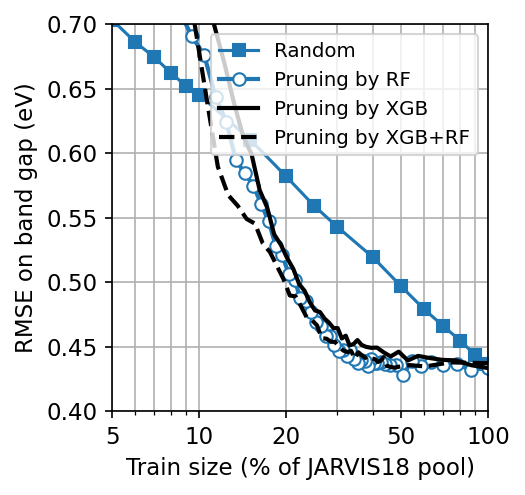}
\includegraphics[width=0.4\linewidth]{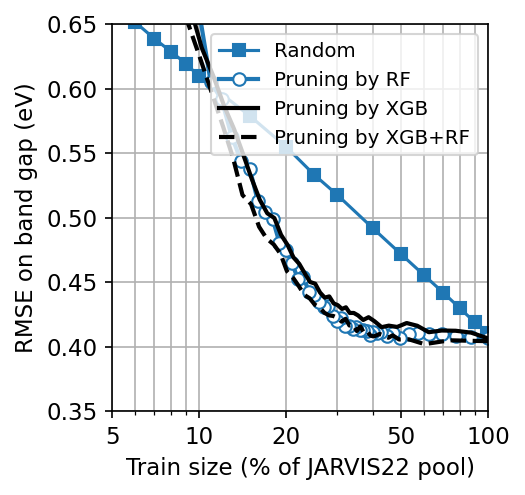}
\includegraphics[width=0.4\linewidth]{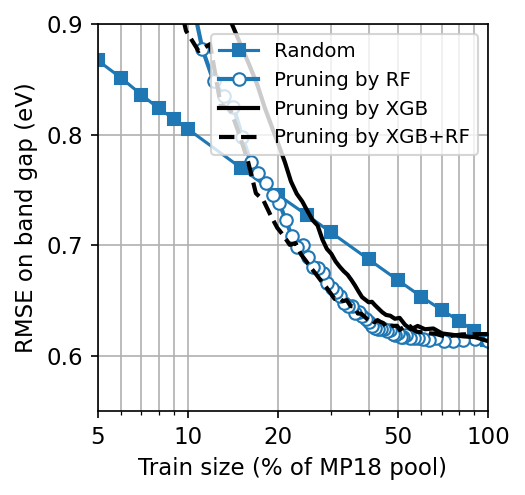}
\includegraphics[width=0.4\linewidth]{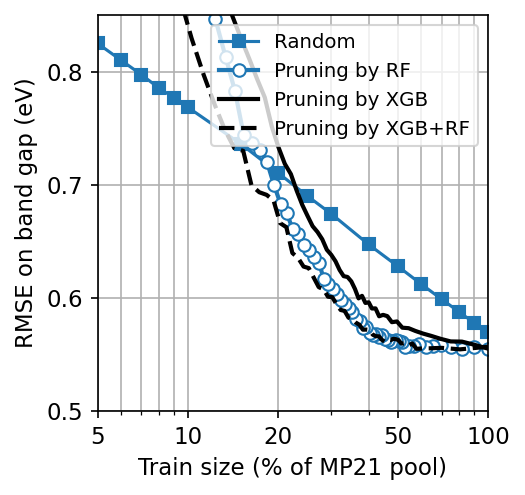}
\includegraphics[width=0.4\linewidth]{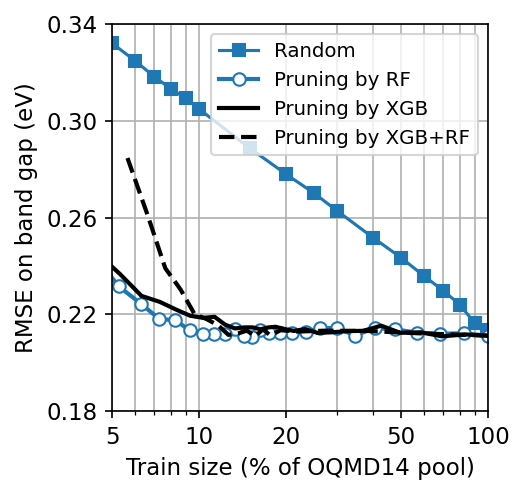}
\includegraphics[width=0.4\linewidth]{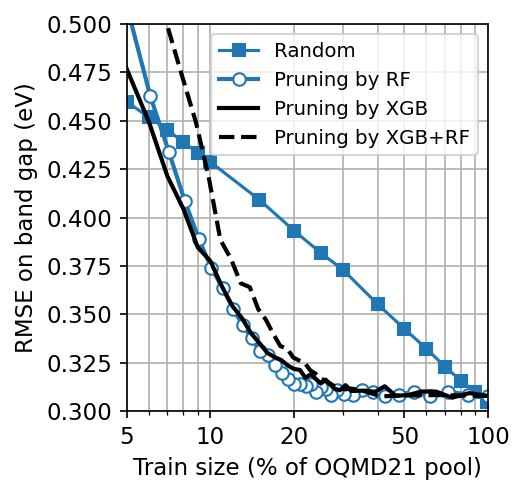}
\caption{\label{fig:trans_ml rf bandgap}
ID performance of the RF models for JARVIS18, JARVIS22, MP18, MP21, OQMD14, and OQMD21 band gap datasets. For each dataset, the RMSE scores obtained by training the RF models on the randomly selected data, the data pruned by the RF models, the data pruned by the XGB models, and the data jointly pruned by the XGB and RF models are shown.
}
\end{figure*}

To investigate the transferability of material sets between ML architectures, we evaluate the ID performance of the XGB and RF models trained on the data pruned by the RF and XGB models, respectively. For the formation energy prediction, the ID performance of the XGB and RF models are shown in Fig.~\ref{fig:trans_ml xgb e_form} and Fig.~\ref{fig:trans_ml rf e_form}, respectively. For the band gap prediction, the ID performance of the XGB and RF models are shown in Fig.~\ref{fig:trans_ml xgb bandgap} and Fig.~\ref{fig:trans_ml rf bandgap}, respectively.

\FloatBarrier
\subsection{Transferability between material properties}

\begin{figure*}[!htbp]
\centering
\includegraphics[width=0.4\linewidth]{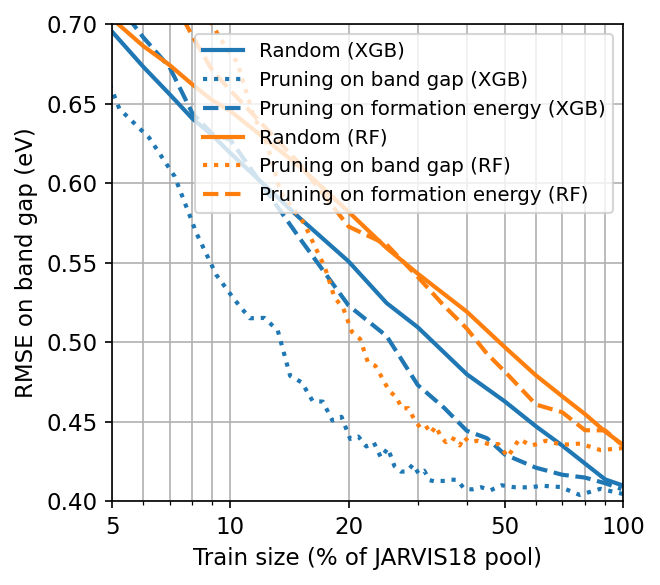}
\includegraphics[width=0.4\linewidth]{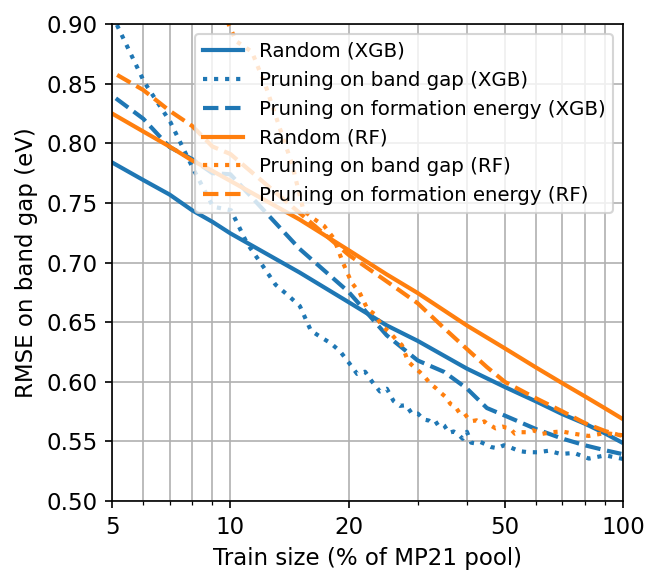}
\includegraphics[width=0.4\linewidth]{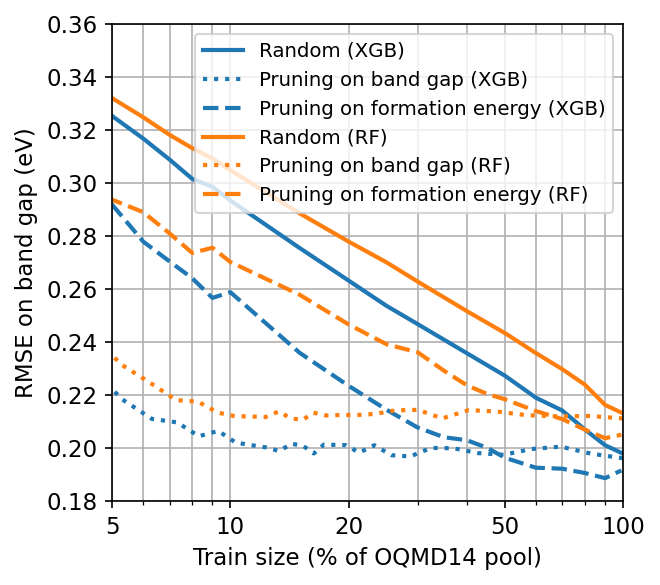}
\caption{\label{fig:trans_prop}
ID performance of the XGB and RF models for JARVIS18, MP21, and OQMD14 band gap datasets. For each dataset and each model, the RMSE scores by training on the randomly selected materials, the materials from the band gap data pruning, and the materials from the formation energy data pruning are shown.
}
\end{figure*}

To investigate the transferability of material sets between material properties, we first identify the material sets from the formation energy data pruning procedure of a given ML model (XGB or RF), and use the corresponding band gap data of these material sets to train the given ML model. The resulting ID performance for JARVIS18, MP21, and OQMD14 band gap datasets is shown in Fig.~\ref{fig:trans_prop}. While the band gap models trained on the data identified from the formation energy data pruning still outperform the models trained on randomly sampled data but by a small degree, and perform less better the models trained on the data from the band gap data pruning, suggesting a limited transferability of the informative material sets between the formation energy and band gap data.

\section{Uncertainty-based active learning\label{sec:AL}}

\begin{figure*}[!htbp]
\centering
\includegraphics[width=0.4\linewidth]{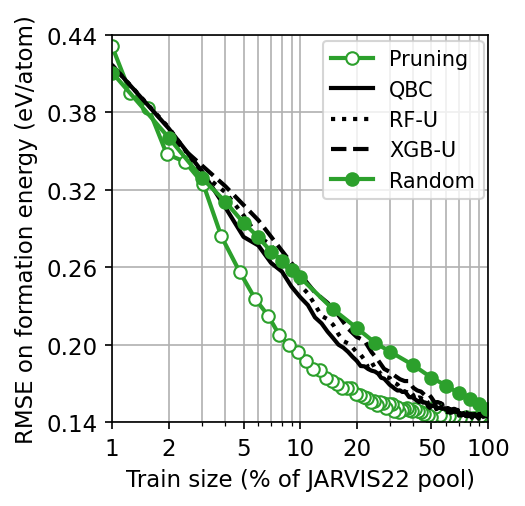}
\includegraphics[width=0.4\linewidth]{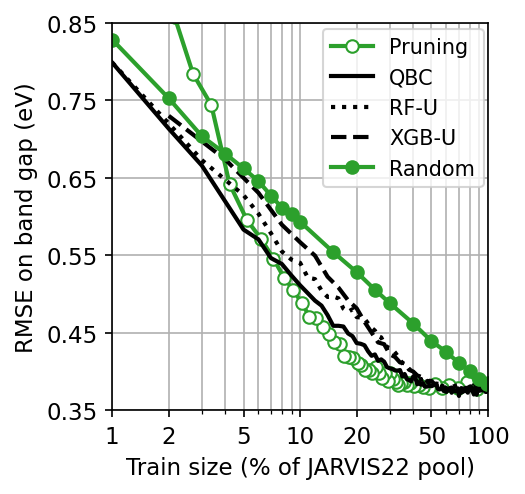}
\includegraphics[width=0.4\linewidth]{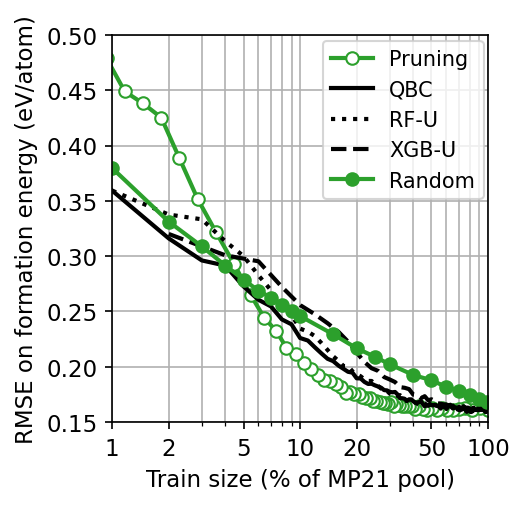}
\includegraphics[width=0.4\linewidth]{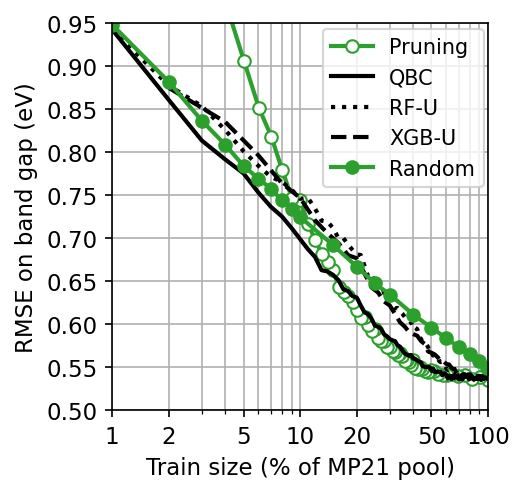}
\includegraphics[width=0.4\linewidth]{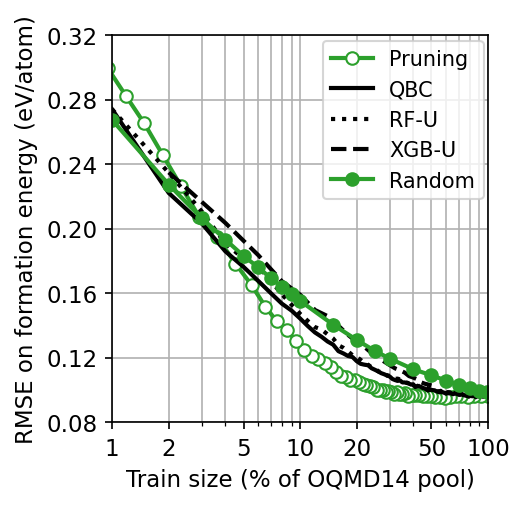}
\includegraphics[width=0.4\linewidth]{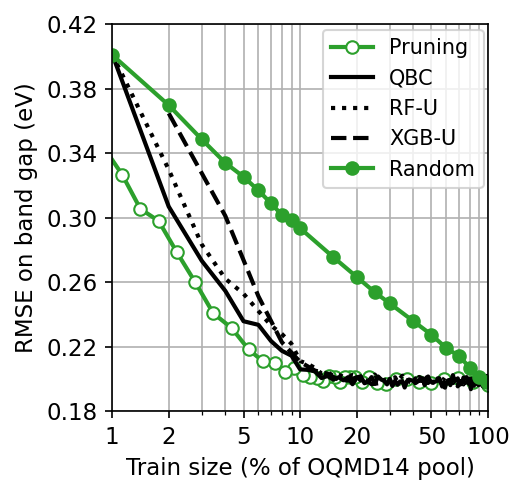}
\caption{\label{fig:al xgb}
ID performance of the XGB models for JARVIS22, MP21, and OQMD14 formation energy and band gap datasets, using the uncertainty-based active learning algorithms. For each dataset, we show the RMSE scores obtained using the randomly selected data, the XGB-pruned data, and the data selected using the uncertainty-based active learning algorithms.
}
\end{figure*}

\begin{figure*}[!htbp]
\centering
\includegraphics[width=0.4\linewidth]{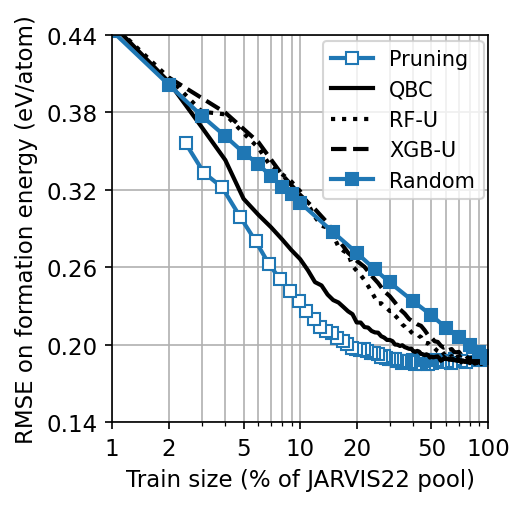}
\includegraphics[width=0.4\linewidth]{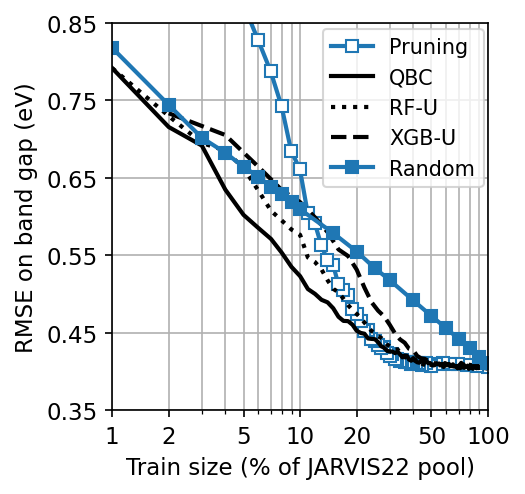}
\includegraphics[width=0.4\linewidth]{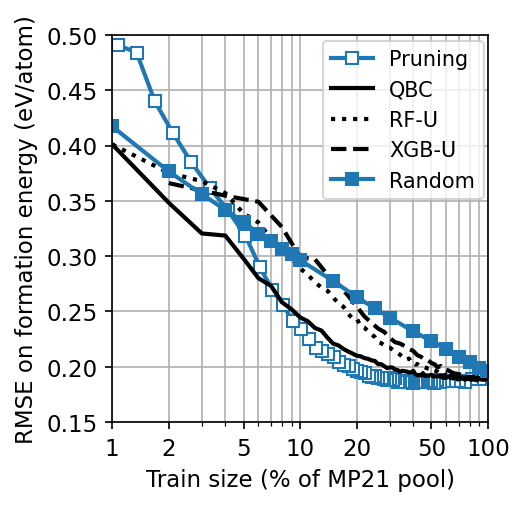}
\includegraphics[width=0.4\linewidth]{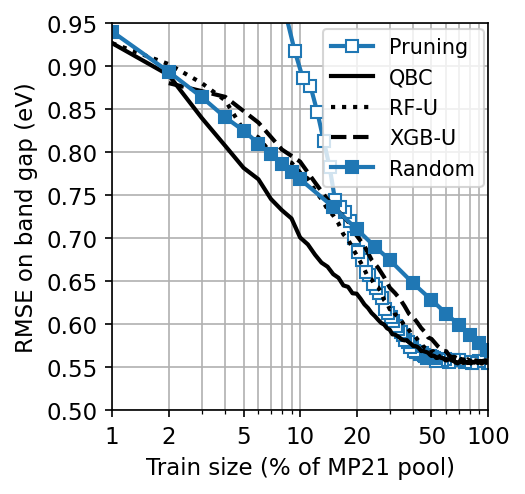}
\includegraphics[width=0.4\linewidth]{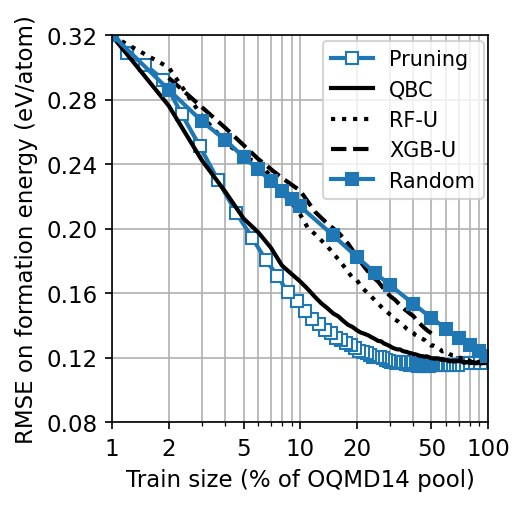}
\includegraphics[width=0.4\linewidth]{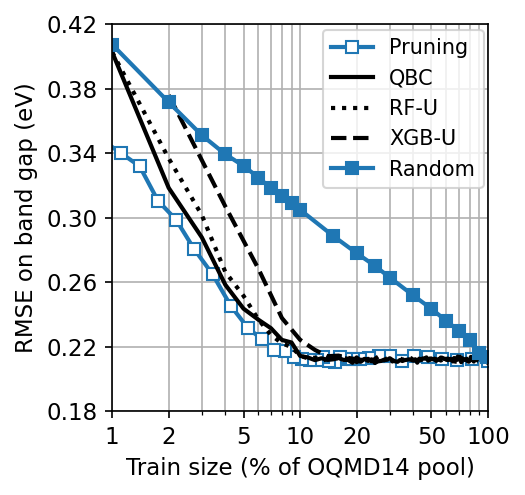}
\caption{\label{fig:al rf}
ID performance of the RF models for JARVIS22, MP21, and OQMD14 formation energy and band gap datasets, using the uncertainty-based active learning algorithms. For each dataset, we show the RMSE scores obtained using the randomly selected data, the RF-pruned data, and the data selected using the uncertainty-based active learning algorithms.
}
\end{figure*}

To demonstrate the feasibility of building smaller but informative datasets, we use uncertainty-based active learning algorithms to grow the JARVIS22, MP21 and OQMD14 datasets from scratch. Three uncertainty measures are considered: The first one (RF-U) is based on the uncertainty of the RF model and is calculated as the difference between the 95th and 5th percentile of the tree predictions in the forest. The second one (XGB-U) is based on the uncertainty of the XGB model using an instance-based uncertainty estimation for gradient-boosted regression trees developed in Ref.~\cite{Brophy2022}. The third one (QBC) is based on the query by committee, where the uncertainty is taken as the difference between the RF and XGB predictions. Fig.~\ref{fig:al xgb} and \ref{fig:al rf} show the resulting ID performance of the XGB and RF models. 

\FloatBarrier
\section{Statistical overlap between the training pool and OOD data\label{sec:umap}}

In Fig. 4 of the main text, we demonstrate that there is a strong OOD performance degradation for the formation energy prediction for the JARVIS and MP datasets, and the band gap prediction for the OQMD dataset. To better understand the statistical overlap between the older dataset and the OOD data in the newer dataset, we provide below an analysis in the feature space using Uniform Manifold Approximation and Projection (UMAP), which is a stochastic and non-linear dimensionality reduction algorithm that preserves the data's local and global structure~\cite{umap}. This technique was previously used to analyze a similar OOD performance degradation for the formation energy prediction of metallic alloys in the MP dataset in our recent work~\cite{li2022critical}, and interested reader is referred to the work therein for an extended discussion. 

Here we take the JARVIS database as an example to describe how the UMAP analysis is done. We first perform the standardization of Matminer-extracted features over the whole JARVIS22 dataset. Then we sequentially drop highly correlated features using a Pearson correlation threshold of 0.7. Next, the remaining features are used as inputs of UMAP to create the embeddings for the whole JARVIS22 dataset. We use n\_neighbors=300 and the default values for other UMAP hyperparameters. We visualize the data in the UMAP embedding with three subplots (see the 1st row in Supplementary Figure~\ref{fig:umap}): the first subplot shows the JARVIS18 data on top of the new data in JARVIS22 (OOD data); the second subplot shows the JARVIS18 only; the third subplot show the OOD data, which are colored by the prediction errors of the XGBoost model trained on the whole JARVIS18 dataset. The results for the MP and OQMD datasets are obtained similarly (see the 2nd and 3rd rows in the 1st row in Supplementary Figure~\ref{fig:umap}).

For the JARVIS database, the OOD data is almost covered by the training data in the UMAP-projected feature space. For the MP database, the OOD data also largely overlap with the training data, with only two small clusters less well covered by the training data. For the OQMD database, there is a relatively large portion of OOD data lying beyond the region covered by the training data. The degree of the overlap might explain to some extent the degree of the OOD performance degradation: for instance, the overlap for the JARVIS database almost reaches 100 \%, which may explain why its OOD performance degradation is the least significant among the three datasets (the Fig. 4 of the main text). However, the subplots showing the OOD data colored by their prediction errors indicate that the OOD data with large prediction errors also occur in the region well covered by the training data. This suggests that more in-depth analysis is needed in the future work to better understand the correlation between the overlap in the feature space and the prediction errors.

\begin{figure*}[!htbp]
\centering
\includegraphics[width=0.8\linewidth]{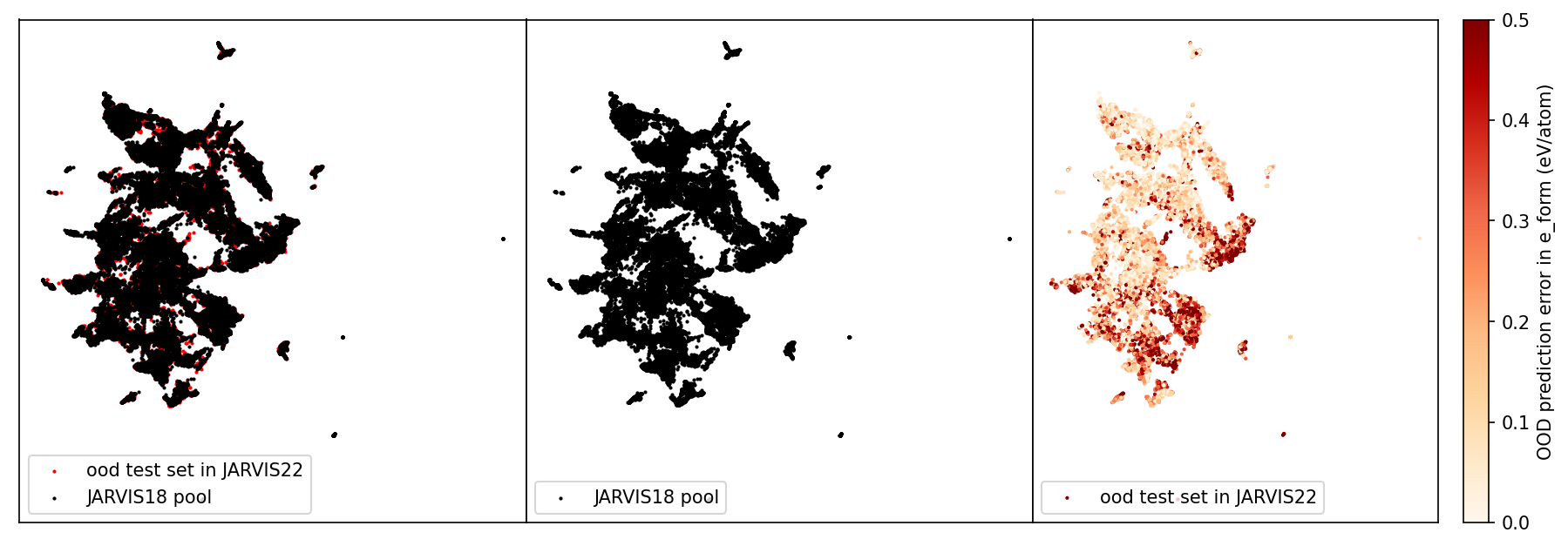}
\includegraphics[width=0.8\linewidth]{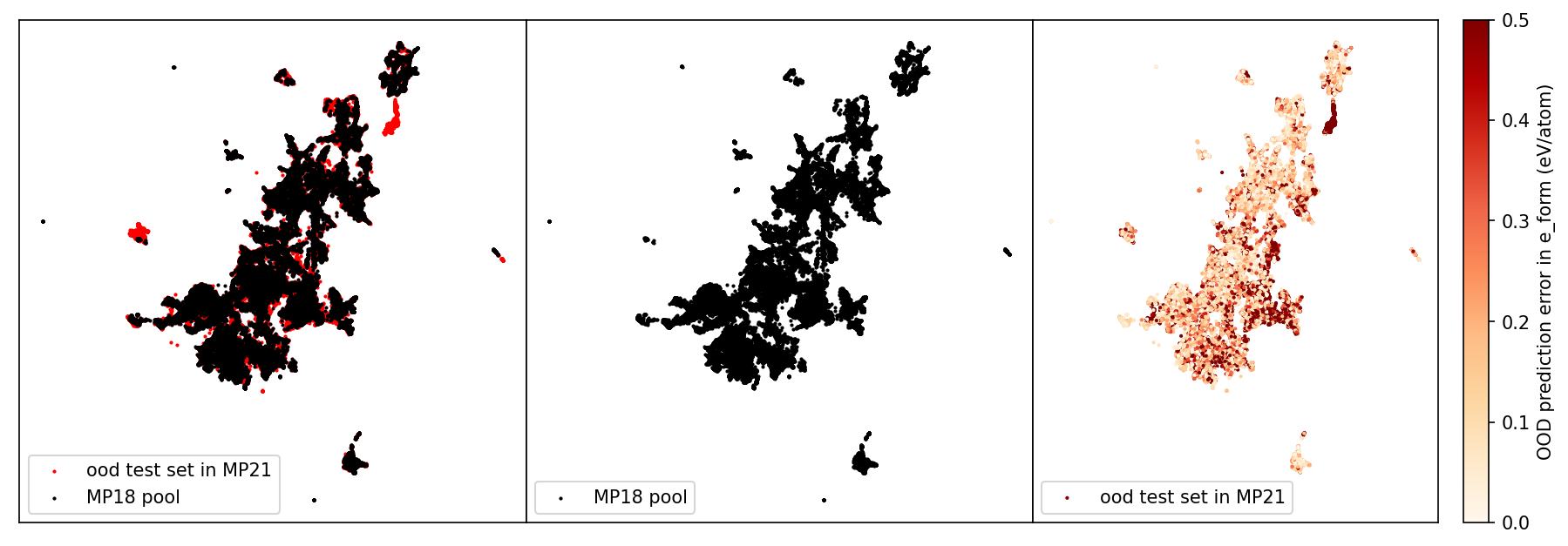}
\includegraphics[width=0.8\linewidth]{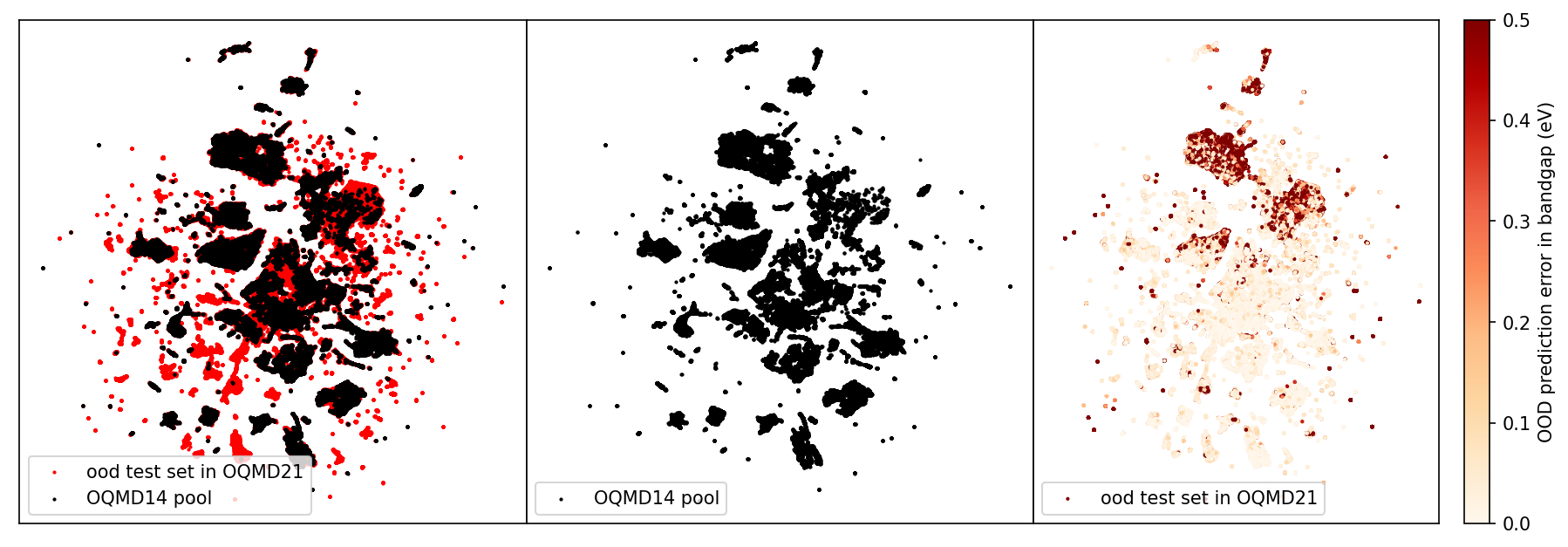}
\caption{\label{fig:umap} UMAP projection in the feature space.
}
\end{figure*}

\bibliography{lib_new}